\documentclass{aa}  

\usepackage[varg]{txfonts}
\usepackage{graphicx, subfigure}
\usepackage{txfonts}
\usepackage{lscape}
\usepackage[usenames,dvipsnames]{xcolor}
\usepackage{hyperref}
\usepackage{multirow}
\usepackage{longtable}
\usepackage{amsmath}

\begin{document} 
\title{Starburst galaxies in the COSMOS field: clumpy star-formation at redshift $0<z<0.5$}

   \author{R. Hinojosa-Go\~ni
          \inst{1,2}
          \and
            C. Mu\~noz-Tu\~n\'on
            \inst{1,2}
            \and
            J. M\'endez-Abreu     
            \inst{3}    
           }

   \institute{Instituto de Astrof\'isica de Canarias, Calle V\'ia L\'actea s/n, E-38200 La Laguna, Tenerife, Spain\\
              \email{hinojosa@iac.es}
         \and Departamento de Astrof\'isica, Universidad de La Laguna,  E-38205 La Laguna, Tenerife, Spain
         \and School of Physics and Astronomy, University of St Andrews, SUPA, North Haugh, KY16 9SS, St Andrews, UK
             }

\date{Received; accepted}
    \abstract
   { At high redshift, starburst galaxies present irregular morphologies, with 10-20\%\,of their star formation occurring in giant clumps. These clumpy galaxies are considered to be the progenitors of local disk galaxies. To understand the properties of starbursts at intermediate and low redshift, it is fundamental to track their evolution and possible link with the systems at higher $z$.}
   {We present an extensive, systematic, and multi-band search and analysis of the starburst galaxies at redshift ($0 < z < 0.5$) in the COSMOS field, as well as detailed characteristics of their star-forming clumps by using Hubble Space Telescope/Advance Camera for Surveys (HST/ACS) images.}
 {The starburst galaxies are identified using a tailor-made intermediate-band color excess selection, tracing the simultaneous presence of H$\alpha$ and [OIII] emission lines in the galaxies. Our methodology uses previous information from the zCOSMOS spectral database to calibrate the color excess as a function of the equivalent width of both spectral lines. This technique allows us to identify 220 starburst galaxies at redshift $0<z<0.5$ using the SUBARU intermediate-band filters.  Combining the high spatial resolution images from the HST/ACS with ground-based multi-wavelength photometry we identify and parametrize the star-forming clumps in every galaxy. Their principal properties, sizes, masses, and star formation rates are provided.}
   { The mass distribution of the starburst galaxies is remarkably similar to that of the whole galaxy sample with a peak around $M/M_{\odot} \sim$ 2$\times$10$^8$ and only a few galaxies with $M/M_{\odot} >$ 10$^{10}$. We classify galaxies in three main types, depending on their HST morphology: Single Knot (Sknot), Single star-forming Knot plus diffuse light (Sknot+diffuse), and Multiple star-forming Knots (Mknots/clumpy) galaxy. We found a fraction of Mknots/clumpy galaxy f$_{clumpy}$=0.24 considering our total sample of starburst galaxies up to $z\sim0.5$.
The individual star-forming knots in our sample follow the same L(H$\alpha$) vs. size scaling relation than local giant HII regions \citep{Fuentes-Masip2000}. However, they slightly differ from the one provided using samples at high redshift. This result highlights the importance of spatially resolving the star-forming regions for this kind of study.
Star-forming clumps in the central regions of Mknots galaxies are more massive, and present higher star formation rates, than those in the outskirts. This trend is smeared when we consider either the mass surface density or surface star formation rate. Sknot galaxies do show properties similar to both dwarf elliptical and irregulars in the surface brightness ($\mu$) versus M$_{host}$ diagram in the $B-$band \citep{amorin2012}, and to  spheroidals and ellipticals in the $\mu$ versus M$_{host}$ diagram in the $V-$band \citep{kormendy2012}.}
   { The properties of our star-forming knots in Sknot+diffuse and Mknots/clumpy galaxies support the predictions of recent numerical simulations claiming that they have been produced by violent disk instabilities. We suggest that the evolution of these knots involves that large and massive clumps at the galaxy centers represent the end product of the coalescence of surviving smaller clumps from the outskirts. Our results support this mechanism and make it unlikely that mergers are the reason behind the observed starburst knots.  
Sknot galaxies might be transitional phases of the BCD class, with their properties consistent with spheroidal like, but blue structures.
   
} 
  
  
\maketitle
\section{Introduction}
\label{sec:intro}

Understanding the physical processes leading to the formation and evolution of galaxies represents a major question of modern astronomy. Among them, the processes governing the star formation (SF) over cosmic time play a fundamental role in galaxy evolution. At high redshift, the SF in galaxies is mainly fueled by accretion of pristine gas from the cosmic web \citep{keres2005,dekel2009a,aumer2010}. In this scenario, when the dark matter halo is diffuse enough, the cool gas stream from the cosmic web can reach the inner halo, or disk, directly providing fresh gas to form stars. This mechanism is called cold-flow accretion, and it is predicted to be the main mode to trigger the SF in the early universe \citep{huillier2012}. Recently, it has been proven that the accretion of metal-poor gas from the cosmic web may also activate the SF in the disk of nearby galaxies \citep{sanchez-almeida2013, sanchez-almeida2014}. Another scenario used to explain the growth of galaxies at high redshift invokes galaxy mergers \citep{conselice2003,bell2006,lotz2006,bournaud2009}. In this case, numerical simulations predict that major mergers can contribute up to 20\% of the galaxy mass growth \citep{wang2011} but at the cost of destroying the galactic disks of their progenitor \citep{naab2006,guo2011}. Therefore, several works suggest that merger-driven starbursts are less important than those triggered by gas accretion from the cosmic web \citep{vandevoort2011}. 

Recent observations and numerical simulations agree, showing that the SF at high redshift occurs mainly in giant clumps \citep{bournaud2015}. The formation mechanism of these SF clumps is still a matter of debate. One scenario proposes that they are formed by disk fragmentation in gravitationally unstable disks \citep{Noguchi1999,Immeli2004a,Immeli2004b,Bournaud2007,Elmegreen2008}. In this case, an intense inflow of cool gas is necessary to provide the high gas surface densities leading to the disk instabilities \citep{dekel2009a}. Contrary to this {\it in-situ} clump formation, \citet{Mandelker2014} proposed that a limited number of {\it ex-situ} clumps might also be accreted by minor mergers into the galaxy disk.

The Hubble Deep Field (HDF) and Ultra Deep Field (UDF) have been crucial to characterize the different morphologies of galaxies at high redshift. Observationally, galaxies at high redshift observed in the HDF or UDF with the Hubble Space Telescope (HST) show clumpy structures with high star formation rates (SFR), which are rare in the local universe \citep{abraham1996,vandenbergh1996,elmegreen2004a,elmegreen2004,Elmegreen2007}. The shape and distribution of the SF knots led \citet{Elmegreen2007} to propose a classification of galaxies as: chains, doubles, tadpoles, clump-clusters, spirals and ellipticals. \citet{elmegreen2004a,elmegreen2004} compared the properties of the clumps present in chain galaxies with those of clump-cluster galaxies. They found similar properties, indicating that both are the same kind of galaxies but seen from different line-of-sights. \citet{genzel2011} studied the properties of five clumpy galaxies at $z$$\sim$2 with deep SINFONI AO spectroscopy. They suggest that the spatial variation of the inferred gas-phase oxygen abundance is broadly consistent with an inside-out growing disk, and/or with inward migration of the clumps. Several studies have confirmed the young nature of the stellar populations in the SF clumps \citep{elmegreen2009,wuyts2012}, with only some rare examples having older stellar populations \citep{bournaud2008}. At low redshift, SF occurs in a large variety of physical scales. Recently, \citet{Elmegreen2013} made use of the Kiso survey of galaxies \citep{miyauchi2010} to compare their SF regions with the giant clumps found at high redshift. They found that local clumpy galaxies in the Kiso survey appear to be intermediate between high-redshift clumps and those in normal spirals. Evidence for SF sustained by tadpole extremely metal poor (XMP) galaxies via cold gas accretion at low redshift has been shown by \citet{sanchez-almeida2014}.

Numerical simulations predict that, for a given mass, the galaxies become more clumpy at high redshift \citep{ceverino2010}. The observations by \citet{Elmegreen2007}, including galaxies of different masses and at different redshifts, confirm this tendency. These SF clumps are important, not only regarding the mass growth of the galaxy, but also for its morphological evolution. Numerical simulations show that SF clumps can migrate from the outer disk to the galaxy center, and contribute to the formation of bulges. The coalescence of clumps can take place in timescales of $\sim$4 rotation times (0.5-0.7 Gyr), therefore representing an alternative path for bulge formation at high redshift \citep{Noguchi1999,Immeli2004a,Bournaud2007,ceverino2010}. The pre-existing thick disk and the SF clump migration to the center have been considered also as responsible for the exponential profile in the luminosity of the present spiral galaxies \citep{Bournaud2007,Elmegreen2013}. As a result of this evolution, the system transforms from an initially uniform disk with high mass SF clumps to spiral-like galaxies with an exponential or double-exponential disk profile, a central bulge, and small remaining clumps \citep{Bournaud2007,ceverino2010,Elmegreen2013}. 

Following the predictions of numerical simulations, observational studies of clumpy galaxies have been performed using mainly high-redshift samples ($z>1$). Deep, high-redshift surveys have proven to be a clue for studying the evolution of disk galaxies, with investigations spanning from the characterization and precise study of the SF knots (clumps), to their relation to the host galaxies. However, these high-redshift studies are usually severely limited by the spatial resolution of the images, even when using space-based observations with the HST Advanced Camera for Surveys (ACS). At lower redshift, studies have been mostly restricted to spectroscopic samples \citep{amorin2012}. We aim to bridge the gap between high-redshift galaxy studies and those in the local universe by providing an accurate characterization of starburst galaxies at intermediate redshift, to constraint models of disk galaxy formation. In the present work we analyze a photometrically selected complete sample of starburst galaxies at redshift $0 < z < 0.5$. The analysis and catalogue of the targets at redshift $z > 0.5$ will be the topic of a forthcoming publication.

The structure of this paper is as follows: In Section \ref{sec:cosmos} we describe the databases used in this study, and our methodology, to identify starburst galaxies in the COSMOS field. In Section \ref{sec:mass} we determine the K-correction and stellar masses of our sample galaxies. In Section \ref{sec:morph} we analyze the morphology of the galaxies, and derive the properties of the starburst knots using the HST images. Section \ref{sec:discussion} presents the properties of the galaxies and star-forming regions. The comparison with high-redshift galaxies is also shown. In Section \ref{sec:results} we provide the conclusions. A standard $\Lambda$CDM cosmology with $\Omega_m$=0.27, $\Omega_{\Lambda}$=0.71, and $H_0$=0.7 is adopted throughout this paper. 

\section{COSMOS-Selected Starburst sample}
\label{sec:cosmos}

COSMOS is the largest deep field survey ever done by the HST. It covers two equatorial square degrees, and about 2$\times$10$^9$ galaxies have been observed in this area with the ACS. Space and ground telescopes have been used to map a wide range of the electromagnetic spectrum, from radio waves to X-ray. In this work we use the photometric redshift catalogue \citep{ilbert2008}, the COSMOS intermediate and broad-band photometric catalogue \citep{capak2007} and  the spectra available in zCOSMOS \citep{lilly2007}, to search for starburst galaxies.

\subsection*{\small {COSMOS Intermediate and Broad Band Photometry Catalogue}}

Over 2 million sources have been observed with the HST/ACS high spatial resolution camera, using the F814W filter in the COSMOS field. Additional observations in 30 bands, covering the ultraviolet (UV) to the infrared (IR), were done using different ground- and space-based instruments, and are available in this catalogue. To perform our search we used the observations with the Intermediate-Band Filters from the SUBARU telescope.

\subsection*{\small {COSMOS Photometric Redshift Catalogue}}

To take spectra of thousands of sources is very time consuming. Therefore, measuring redshifts based only on photometric data is more efficient, and has been demonstrated to be a powerful tool in recent surveys \citep{ilbert2009,benitez2009}. The photometric redshift ($z_{phot}$) in COSMOS was computed using 30 broad, intermediate and narrow-band filters covering the UV, visible-near infrared (NIR) and mid-IR spectral ranges \citep{ilbert2009}. This catalogue contains 385065 sources classified as galaxies, stars, X-ray sources, faint sources or masked areas.   Among them, we choose only the 305002 sources flagged as galaxies. To measure the H$\alpha$ and [OIII] emission lines we used the SUBARU filters and the $z_{phot}$. To this aim, we matched the COSMOS Intermediate and Broad-Band Photometry Catalogue with the  COSMOS photometric redshift catalogue.

\subsection*{\small {zCOSMOS}}

The spectroscopic redshift ($z$) was obtained from observations of the VIsible MultiObject Spectrograph (VIMOS)  on the Very Large Telescope (VLT) for a subsample of the COSMOS field, as part of the zCOSMOS project \citep{lilly2007}. This catalogue contains two parts. The first is zCOSMOS-bright, aimed at observing $\sim$20000 galaxies at 0.1 $<  z <$ 1.2 using the VIMOS red spectral range (5550-9650 \AA) with the R $\sim$ 600 MR grism, in order to detect the strong spectral features around 4000 \AA. The second part is the zCOSMOS-deep, which aims at observing $\sim$10000 galaxies lying at 1.5 $< z <$ 2.5 in the blue spectral range (3600-6800 \AA), to measure the strong absorption and emission features in the range between 1200 and 1700 \AA. At the moment, there are 10643 published spectra, corresponding to the results of the zCOSMOS-bright spectroscopic observations, which were carried out in VLT Service Mode during the period April 2005 to June 2006. These have been observed using a 1 arcsec-wide slit, sampling roughly 2.5 \AA/pixel, with a velocity accuracy of $\sim$100 km s$^{-1}$.

\subsection{Starbursts in zCOSMOS}

Starburst galaxies show a steep rising continuum in the blue region of the spectra, combined with strong nebular emission lines. At visible wavelengths, the [OIII] and H$\alpha$ emission lines are the most prominent, and we can parametrize them using the equivalent width (EW). Previous studies of starburst galaxies in the local Universe \citep{kniasev2004,cairos2007,cairos2009a,cairos2009b,cairos2010,morales2011,amorin2014} found minimum values of the [OIII] and H$\alpha$ EW of about 80 \AA. This value corresponds to young star-forming regions, with ages < 10 Myr \citep{leitherer1999,vazquez2005,leitherer2010}.  Therefore, the EW turn out to be the best parameter to identify young starburst galaxies and we adopt the aforementioned EW threshold values in our work.

To measure the EW of the [OIII] and H$\alpha$ emission lines we first used the spectra from zCOSMOS. The wavelength range of the spectra allows us simultaneously to measure [OIII] and H$\alpha$ in the redshift range 0.1 $\le z \le$ 0.47. zCOSMOS comprises 3384 galaxies in this redshift range. To identify those with emission lines, we used the published spectroscopic redshift and the rest-frame wavelengths of the [OIII] and H$\alpha$ emission lines ($\lambda_c$: 4959, 5007 \AA; 6563 \AA). Then, we measure the flux of the emission line (F$_l$) in a baseline of 30 \AA, which allows us to cover the total width of the line at the continuum level. The same band-width was used to measure the flux of the continua (F$_c$) at rest-frame wavelengths 5053 \AA\ and 6518 \AA\  for [OIII] and H${\alpha}$, respectively. The EW is defined as

\begin{equation}
\mathrm{EW}=\sum \frac{F_\mathrm{l}-F_\mathrm{c}}{F_\mathrm{c}}\delta\lambda,
\end{equation}
where $\delta\lambda$=2.5 \AA. The associated error is given by

\begin{equation}
e_\mathrm{EW}=\frac{F_\mathrm{l}}{F_\mathrm{c}} \Bigg(\frac{\sigma_c}{med_c} \cdot \sqrt{8 n} \Bigg) ,\ 
\end{equation}
where both $\sigma_c$ and  $med_c$ correspond to the standard deviation and the median in the continuum respectively, and $n$ is the number of resolution elements in the selected baseline. We calculated the EW in H$\alpha$ and [OIII] for all 3384 galaxies, covering the range $0.1 \le z \le 0.47$.  We selected those with EW $\ge$ 80\AA\ in H${\alpha}$ and [OIII], obtaining a sample of 82 starburst galaxies. Table \ref{tabla_zcosmos} shows the first few entries of the spectroscopic emission-line catalogue. The complete catalogue is available in table 3. 

\begin{table}
\caption{Starburst galaxies selected from zCOSMOS.}
\label{tabla_zcosmos}
\centering
\begin{tabular}{c c c c} \hline\hline       
object & $z$ & EW(H$\alpha$) & EW([OIII]) \\ 
       &          &    (\AA)            &  (\AA)    \\
  (1)  &    (2)   &    (3)              & (4)       \\
\hline
cosmos-002 & 0.34 & 234  $\pm$ 43 & 321  $\pm$ 18 \\
cosmos-003 & 0.25 & 104  $\pm$ 14 & 108 $\pm$ 28\\ 
cosmos-010 & 0.13 & 144 $\pm$ 19 & 100 $\pm$ 11\\ 
cosmos-014 & 0.38 & 208 $\pm$ 21 & 124 $\pm$ 7\\ 
cosmos-015 & 0.41 & 94 $\pm$ 21 & 166 $\pm$ 16\\ 
cosmos-017 & 0.17 & 177 $\pm$ 58 & 121 $\pm$ 12\\ 
cosmos-018 & 0.28 & 193 $\pm$ 40 & 174 $\pm$ 10\\ 
cosmos-021 & 0.19 & 96 $\pm$ 19 & 126 $\pm$ 8\\ 
cosmos-026 & 0.44 & 148  $\pm$ 70 & 115 $\pm$ 18\\ 
cosmos-027 & 0.34 & 190 $\pm$ 31 & 164 $\pm$ 6\\ \hline
\end{tabular}
\tablefoot{(1) Object name (ordered by RA), (2) spectroscopic redshift, (3) H${\alpha}$ EW and (4) [OIII] EW.}
\end{table}

\subsection{Starbursts in the COSMOS Photometric Catalogue}

Only $\sim$2.8\% of objects in COSMOS have a spectroscopic redshift. Therefore, the use of the photometric redshift is fundamental to produce a complete catalogue of starburst galaxies.  In the following, we will describe the color-color selection method used to detect galaxies with EW $\ge$ 80\AA\ in both H${\alpha}$ and [OIII]. We also show how we calibrate the method using the available spectral information.

\subsubsection{Calibration of color-color diagnostics with spectra}

Color-color diagrams using narrow/intermediate bands have been successfully used in the past to select samples of emission line-galaxies \citep{sobral2013}. However, the definition of the boundaries between different galaxy types, generally star forming versus passive, is often subjective, and dependent on the characteristics of the objects under study. 

\begin{figure}[!t]
\centering
\hspace{-1cm}
\includegraphics[width=9.9cm]{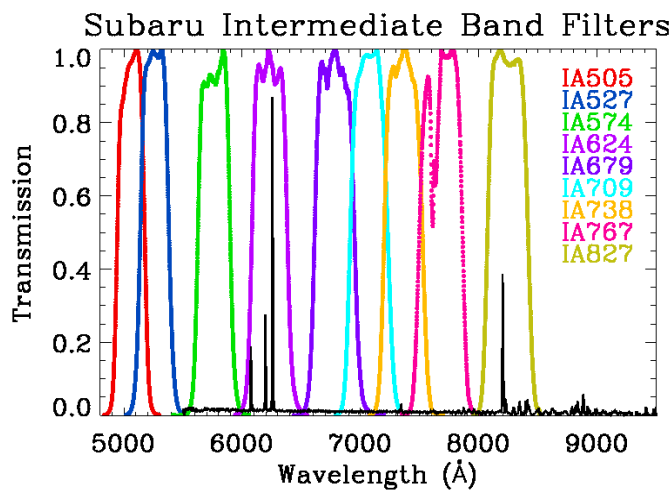}
\caption{{\small Normalized transmission curves of the intermediate-band SUBARU filters used in this study. A representative emission-line galaxy at $z=0.25$ is also shown to demonstrate how the filters match simultaneously the positions of the [OIII] and H$\alpha$ lines. The space between filters determine the wavelength gaps in our photometric sample.}}
\label{filtros}
\end{figure}

\begin{figure*}[tH]
\centering
\includegraphics[width=15cm]{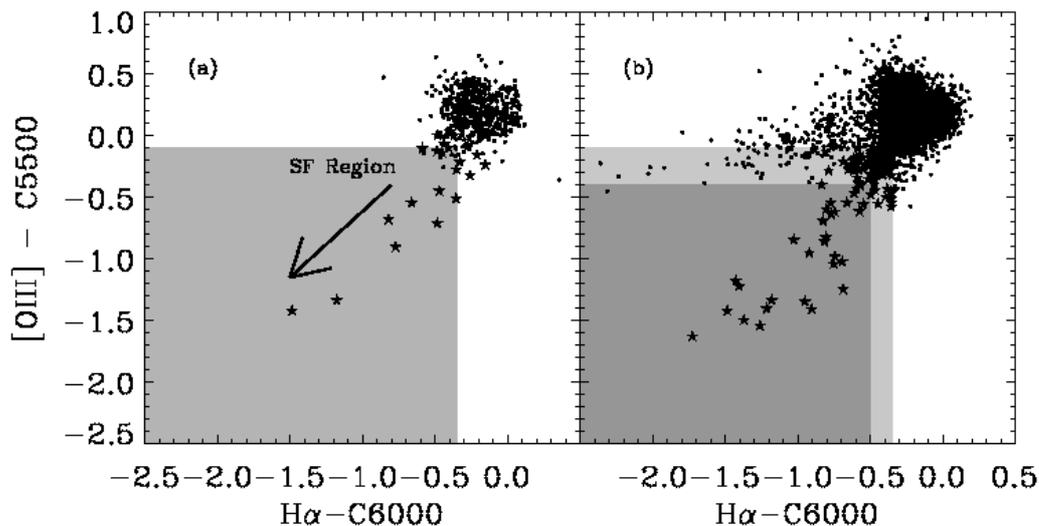}
\vspace{-7.5cm}
\caption{{\small a) Color-color diagram of the 580 galaxies in zCOSMOS measured with SUBARU filters. Star symbols show the galaxies with EW $\ge$ 80 \AA\ in both H${\alpha}$ and [OIII]. The arrow points in the direction of galaxies with larger EW in both emission lines. The shadowed region in the diagram is the starburst location. b) Color-color diagram for galaxies in the photometric redshift catalogue, in the redshift ranges: $0.007\le z \le 0.074$, $0.124 \le z \le 0.177$, and $0.230 \le z \le 0.274$. Star symbols show the galaxies in the star-forming region defined in panel (a). We discarded galaxies with high H$\alpha$ and low [OIII] emission. A subregion (darker shadow) shows the loci in which the photometric EW in H$\alpha$ and [OIII] is larger than 80 \AA\ for the photometric sample.}}
\label{corte_color}
\end{figure*}

In order to calibrate our color-color diagnostic diagram we used the sample of confirmed starburst galaxies in the spectroscopic catalogue (see Section 2.1) as templates to search for a complete sample in the COSMOS photometric sample. These galaxies were selected as starbursts based on their EW in both H${\alpha}$ and [OIII]. We used the spectroscopic redshifts to locate the emission lines in the SUBARU intermediate band filters. We also selected two regions free of lines for the continuum, centered at 5500 \AA\ (C5500) as continuum for the [OIII] emission line, and at 6000 \AA\ (C6000) for the H$\alpha$ emission line. With these we constructed color-color diagrams, matching the presence of the corresponding strong line and its continuum. The colors constructed are SUBARU (H$\alpha$)-SUBARU (C6000) vs. SUBARU ([OIII])-SUBARU (C5500). Fig. \ref{filtros} shows the transmission of the SUBARU filters, with an example starburst galaxy at redshift 0.25. Note that SUBARU filters do not homogeneously cover the whole range; instead some gaps can be clearly seen in Fig. \ref{filtros}. The redshift range where we are able to locate simultaneously H$\alpha$ and [OIII] lines are: $0.123 < z < 0.178$ and $0.23 < z < 0.274$. The total number of galaxies in zCOSMOS, within these redshifts, are 580, of which 24 are starburst galaxies. Their color-color diagram (SUBARU (H$\alpha$)-SUBARU (C6000) versus SUBARU ([OIII])-SUBARU (C5500)) is shown in Fig. \ref{corte_color}a. Galaxies with EW $\ge$ 80 \AA, in both H$\alpha$ and [OIII], measured from the spectra, are represented by stars. The arrow in Fig. \ref{corte_color}a shows the direction of increasing emission-line EW. As expected, starburst galaxies populate a well-defined region of the color-color diagram. The use of galaxies with spectra allow us to calibrate the color excess as a function of the emission-line EWs and therefore using the easily accessible photometric data to select starburst galaxies with EW $\ge$ 80 \AA, in both H$\alpha$ and [OIII].

We use the region where H$\alpha$-C6000 $\le$ -0.35 and [OIII]-C5500 $\le$ -0.1 (shadow region in Fig. \ref{corte_color}a) as the proxy for emission associated with starburst galaxies. This region is used to search for starburst candidates using the photometric redshift catalogue in the next section.  

\subsubsection{Starbursts using the COSMOS photometric redshift catalogue}

The photometric redshift allows us to match the H$\alpha$ and [OIII] emission lines in the SUBARU intermediate-band filters in the redshift ranges: 0.007 $\le z \le$ 0.074, 0.124 $\le z \le$ 0.177, and 0.230 $\le z \le$ 0.274. The bands, including [OIII], H$\alpha$, and their respective continuum at 5500 \AA\ and 6000 \AA, have been used to build a color-color diagnostic diagram (shown in Fig. 2b)  similar to the one for the spectroscopic sample. We limit our sample to galaxies with m(F814W)$<$23.5. Fainter galaxies have both photometric redshift and intermediate-band magnitude errors too large for our analysis. In order to estimate the photometric EW we used a filter in the red continuum for [OIII] and in the blue continuum for H$\alpha$. Then we used Eq. 1 to calculate the EW for both emission lines. In Fig. 2b we plot in the color-color diagram defined using the SUBARU filters for all the galaxies in the zCOSMOS catalogue. The star symbols show the locii for the starburst galaxies within the region with EW > 80 \AA.

 In Fig. \ref{corte_color}b a smaller region is shown with a darker shadow, with H$\alpha$-C6000 $\le$ -0.5 and [OIII]-C5500 $\le$ -0.4. In the larger region, which we used to obtain our sample, only 33\% of the photometric EW in H$\alpha$ or [OIII] are less than 80 \AA. The smaller region exclusively comprise objects where the photometric EW in both H$\alpha$ and [OIII] is > 80 \AA. These are the targets included in the catalogue presented in this work. Table \ref{tabla_subaru} shows some of the entries of the photometric emission-line catalogue, constructed using only the galaxies in the darker shadowed area (star symbols in Fig. \ref{corte_color}b). The complete catalogue is available in table 3.
    
\begin{table}
\caption{Starburst galaxies selected from the photometric catalogue.}
\label{tabla_subaru}
\centering
\begin{tabular}{c c c c}  
\hline
\hline
Object & $z_{phot}$ & EW(H$\alpha$) & EW([OIII])\\ 
       &          &    (\AA)            &  (\AA)    \\
  (1)  &    (2)   &    (3)              & (4)       \\
\hline
cosmos-001 & 0.27 & 760 $\pm$ 46 & 770 $\pm$ 43 \\ 
cosmos-004 & 0.26 & 119 $\pm$ 26 & 171 $\pm$ 26 \\ 
cosmos-005 & 0.26 & 94  $\pm$ 10 & 182 $\pm$ 11 \\ 
cosmos-006 & 0.26 & 124 $\pm$ 11 & 125 $\pm$ 10 \\ 
cosmos-007 & 0.26 & 108 $\pm$ 19 & 112 $\pm$ 16 \\ 
cosmos-008 & 0.16 & 110 $\pm$ 34 & 60 $\pm$  30 \\ 
cosmos-009 & 0.06 & 28 $\pm$ 10 & 109 $\pm$ 11 \\ 
cosmos-011 & 0.26 & 351 $\pm$ 13 & 568 $\pm$ 16 \\ 
cosmos-012 & 0.01 & 114 $\pm$ 7 & 55 $\pm$  5 \\ 
cosmos-013 & 0.26 & 114 $\pm$ 27 & 181 $\pm$ 26 \\ \hline
\end{tabular}
\tablefoot{(1) Object name (ordered by RA), (2) photometric redshift, (3) H${\alpha}$ EW and (4) [OIII] EW.}
\end{table} 

\subsection{Caveats to the sample selection}

Several comparisons have been made to assure that our photometric sample is reliable, and to establish possible sources of uncertainty. We have detected inconsistencies in some of the photometric redshift estimations. In order to have a secure sample, other criteria have been used to filter our sample; galaxies with H$\alpha$ emission-line only (without [OIII]), have been identified in Fig. \ref{corte_color}b (see horizontal branch with 0$<$[OIII]-C5500$<$-0.5) and discarded. Comparing photometric and spectroscopic redshifts we have detected some inconsistencies in the photometric redshift estimation for $z$ $\leq$ 0.1. As the colors shown in the diagram use the photometric redshift, the diagram is also useful to detect inconsistencies in $z_{phot}$ for emission-line galaxies. For these galaxies at $z$ $\leq$ 0.1, the photometric redshift confused [OII] with [OIII], and [OIII] with H$\alpha$ emission lines. This problem, known as "catastrophic redshift", was reported in \citet[][their Fig. 1]{ilbert2008}.  In summary, since we look for starburst galaxies showing simultaneously H$\alpha$ and [OIII], when discarding galaxies in the color-color diagram with high H$\alpha$ and low [OIII], we automatically remove galaxies with wrong photometric redshift determinations.

\begin{figure}[htbp]
\centering
\includegraphics[width=09.0cm]{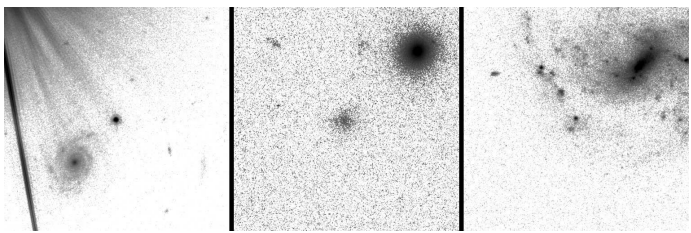}
\caption{{\small HST images with some examples of sources discarded after visual inspection. Left: saturated by a close star. Middle: galaxies at the limit of our detection threshold. Right: HII region of a foreground spiral galaxy.}}
\label{mosaico_01}
\end{figure}

\begin{figure}[htbp]
\centering
\includegraphics[width=09.0cm]{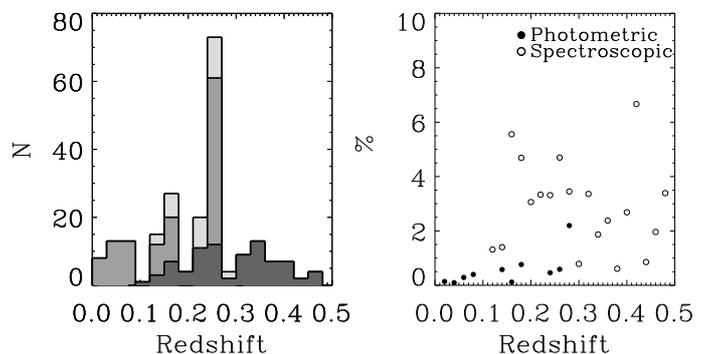}
\caption{{\small Left: Redshift distribution of the photometric (grey), spectroscopic (dark grey) and total (light grey) sample. Right: Percentage of starburst galaxies found in COSMOS, with respect to the total analyzed galaxies in different redshifts. The photometric sample is showed in filled circles, and the spectroscopic sample in open circles.}}
\label{hist_red_per}
\end{figure}

Altogether, in COSMOS we have identified a total number of 289 starburst galaxies with EW $\ge$ 80 \AA\  in H${\alpha}$ and [OIII], either from the photometric or spectroscopic catalogues. From this sample, a total of  69 objects were rejected after a careful visual inspection. Fig. \ref{mosaico_01} shows some examples of fake detections, galaxies  saturated by a close star, galaxies at the limit of our detection threshold, and HII regions of foreground spiral galaxies that were removed from the final sample. Our final catalogue contains 220 starburst galaxies. In Fig. \ref{hist_red_per} (left) we show the distribution in redshift for our sample including the gaps in redshift for the photometric sample. In  Fig. \ref{hist_red_per} (right) the percentage of photometric and spectroscopic sample, with respect to the total number of objects in COSMOS, is shown. As it can be seen, the distribution in redshift in our sample is not homogeneous (as a consequence, our sample distribution has not a perfect completitude).

\section{K-correction and stellar mass determination}
\label{sec:mass}

In this section we compute the stellar mass for the 220 starburst galaxies of the sample. In order to account for redshift effects in the flux measurements for every filter of the photometric catalogue, we first calculated the K-correction for each galaxy. We used the last version of K-correct\footnote{http://howdy.physics.nyu.edu/index.php/Kcorrect} \citep{blanton2007}. This software uses a basis set of 485 spectral templates,  in which 450 of these are a set of instantaneous bursts from \citet{bruzual2003} models, using the \citet{chabrier2003} stellar initial mass function (IMF) and the Padova 1994 isochrones \citep{alongi1991}. The remaining 35 templates are from MAPPINGS-III \citep{kewley2001}, which are models of emission from ionized gas, a crucial feature appearing in the galaxies in our sample. All our galaxies are in the photometric redshift catalogue, and have been mapped by COSMOS using 10 bands ($U, B, V, g, R, I, $ F814W, $z, J,$ and $K$). In some cases, the magnitude in one of the bands was not available ($\sim$10$\%$; the $K-$band usually). In those cases the K-correction was calculated using 9 bands only. In our catalogue (Table 3), every magnitude and color are in rest-frame. 
We have visually analyzed every Spectral Energy Distribution (SED) of the sample to check that they are blue Emission-Line Galaxies. To estimate the goodness of the fit, we have defined the "goodness parameter" (G), which is expressed as $\sum(F_t-F_m)^2/F_t^2$, where $F_t$ is the flux of the template at the wavelength of the filter, and $F_m$ is the measured flux in the filter from COSMOS. The G parameter is also given in Table 3. The best fits from K-correct show some features, such as blue galaxies Balmer break in the continuum and strong emission lines. Two (among the 220) galaxies do not have those features, and the fit is very poor (G very high). They appear with a value of 99 in Table 3.

\begin{figure*}[!ht]
\begin{center}
\includegraphics[width=0.32\textwidth]{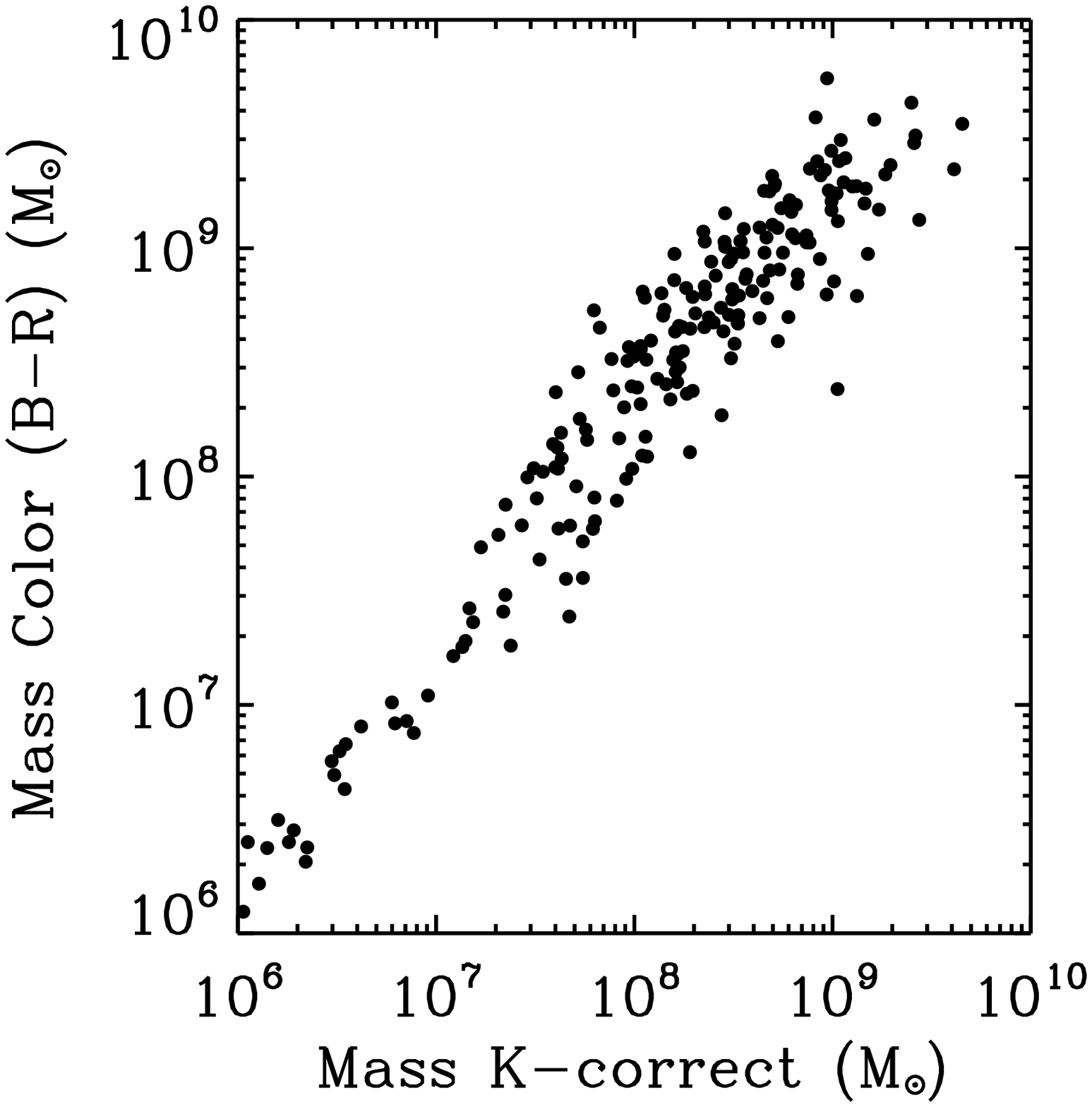}
\includegraphics[width=0.32\textwidth]{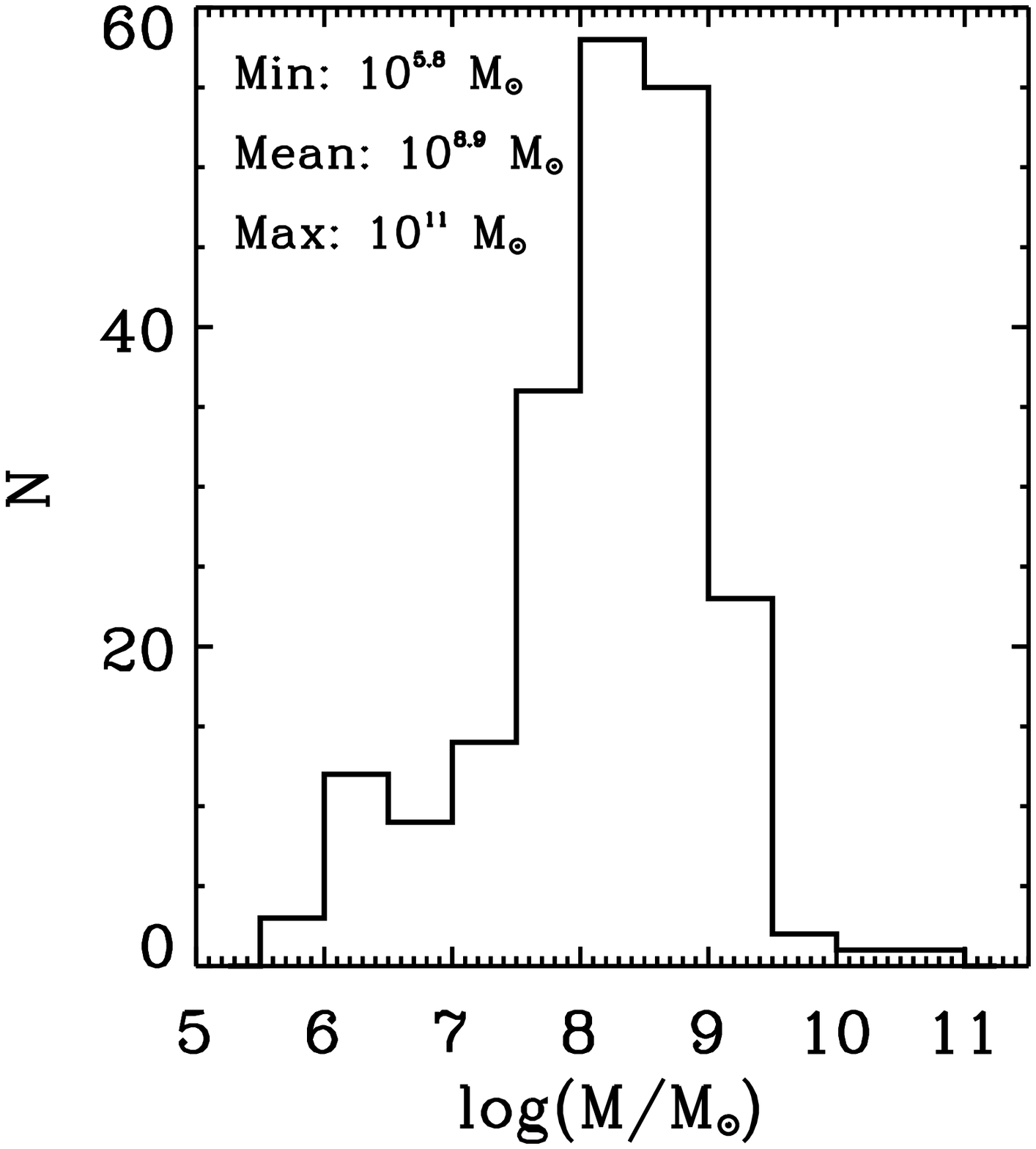}
\includegraphics[width=0.32\textwidth]{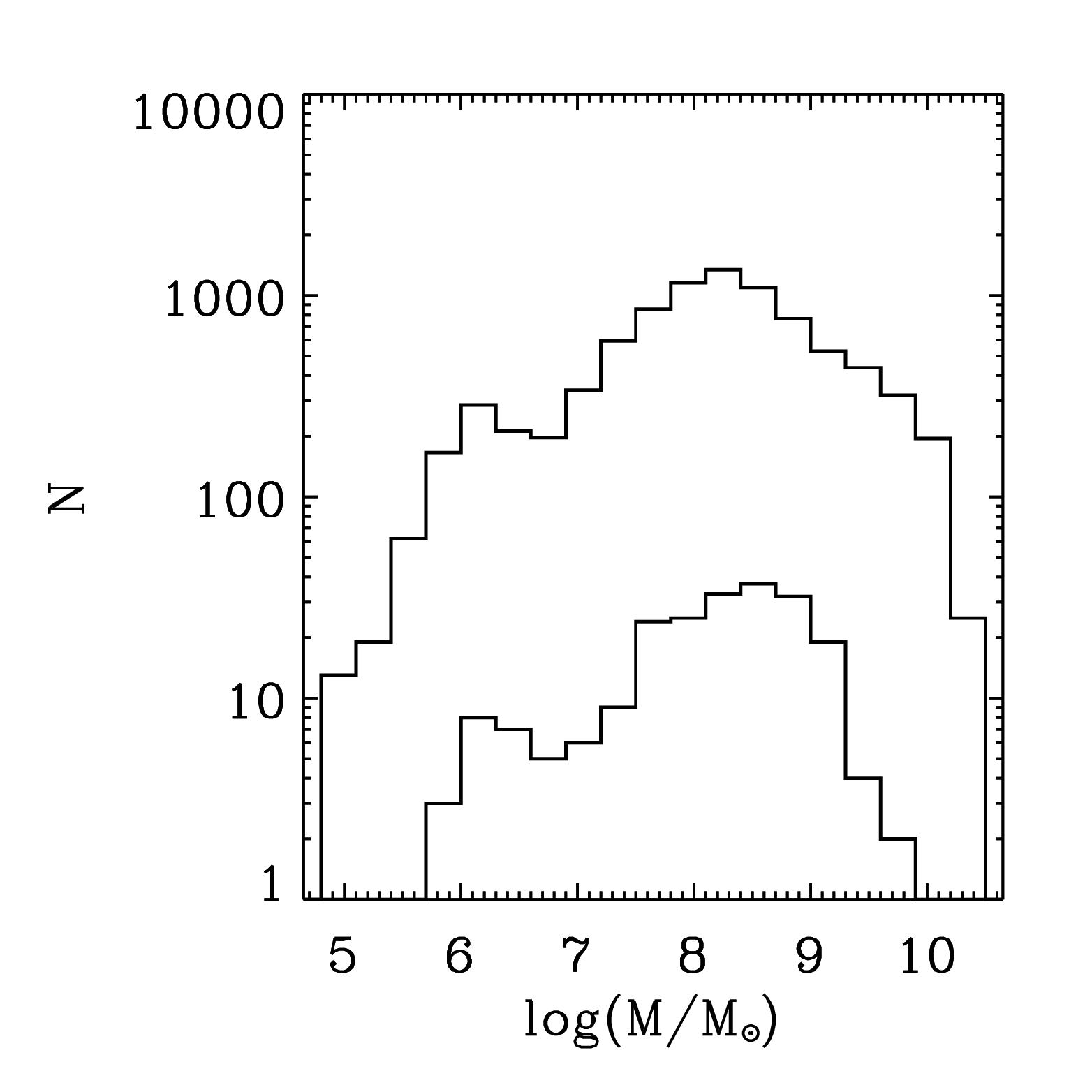}
\end{center}
\vspace{-0.5cm}
\caption{Left panel: Comparison of stellar masses estimated from the K-correct software fitting the galaxy SED, and the $B-R$ color using the prescription by \citet{bell2001}. Middle panel: Stellar mass distribution of the starburst galaxy sample, determined with K-correct. Right panel: Stellar mass distribution for both the entire COSMOS galaxy sample with $z<0.3$ and m(F814W)$<$23.5 and the starburst identified in our paper (in logarithmic scale).}
\label{figmasas}
\end{figure*}

The K-correct software also provides information about the stellar masses of galaxies. These are published in Table 3, and are in the range 10$^{5.8}$ < $M/M_{\odot}$ < 10$^{11}$. It is worth noticing that our selection criteria does not introduce any bias on the mass distribution. The low-mass end of our mass distribution is given by the observational limits (mF814W$ < $23.5) whereas the limited volume probed by the COSMOS survey produces the high-mass end. For the sake of comparison, we also computed stellar masses using the equations provided by \citet{bell2001}. They use a suite of simplified spectrophotometric spiral galaxies, with SF burst, to calculate the stellar mass to luminosity ratio using colors and assuming a universal scaled Salpeter IMF. To apply these equations we used independents three colors, with magnitudes previously K-corrected, to calculate the stellar mass and to check the robustness of our results. The results are consistent with the mass obtained with K-correct. Fig. \ref{figmasas} (left panel) shows the comparison between the stellar mass from K-correct and \citet{bell2001}; both methods to calculate the galaxy mass are consistent. 
Fig. \ref{figmasas} (middle panel) shows the mass distribution of the starburst galaxies determined with K-correct. Most of them are in the 10$^8$-10$^9$ M$_\sun$ range. For comparison, we also computed the stellar masses of the whole COSMOS sample with $z<0.3$ and m(F814W)$<$23.5 (8650 galaxies) using the K-correct algorithm. The result is shown in Fig. \ref{figmasas} (right panel) were both distributions are plot. It is clear that the number of galaxies in the COSMOS volume at this redshift range peaks at $M/M_{\odot} \sim$ 2$\times$10$^8$ with only a few galaxies with $M/M_{\odot} >$ 10$^{10}$. The shape of the mass distribution is actually very similar to that of the starburst galaxies with maybe an excess of galaxies in the high-mass end. We suggest that this excess could be related to more red, elliptical-like galaxies dominating this region of the mass function.

\section{Morphology of the Starburst galaxies}
\label{sec:morph}

\begin{figure}[htbp]
\centering
\includegraphics[width=09.0cm]{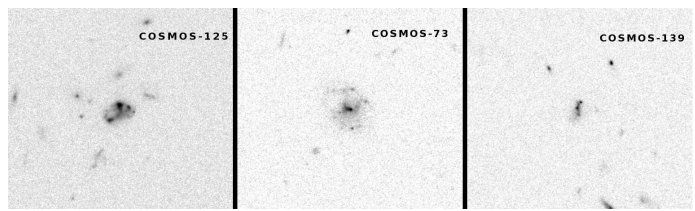}
\caption{{\small F814W HST/ACS 15"$\times$15" high resolution images of three identified Starburst galaxies.}}
\label{mosaico_02}
\end{figure}

The COSMOS field has been targeted by the HST/ACS camera with the F814W filter \citep{capak2007}. This filter is centered at $\lambda_c$=8037 \AA, and its wavelength width is $\Delta\lambda$=1862 \AA.  
The F814W HST/ACS high resolution images of  the sample (220 galaxies), with a Full Width at Half Maximum (FWHM) of the Point Spread Function (PSF) of 0.09" and a pixel scale of 0.03"/pixel, are available in the IRSA's General Catalogue Search Service \footnote{http://irsa.ipac.caltech.edu/data/COSMOS/index\_cutouts.html}, and have been analyzed to identify morphological structures. A semi-automatic protocol was designed to this aim. We downloaded the images with a size of 15"$\times$15" centered in our targets coordinates.  Most galaxies are smaller than this size (see Fig. \ref{mosaico_02} for an example). However, we found that 26  are larger than the downloaded images, and they were analyzed individually. We use Source-Extractor\footnote{http://www.astromatic.net/software/sextractor} \citep{sextractor1996} to obtain the coordinates of the galaxies and put them in the center of the images. To perform a detailed analysis, particularly to identify sub-structures like star-forming regions in starburst galaxies, we used the Faint Object Classification and Analysis System (FOCAS\footnote{http://iraf.noao.edu/ftp/docs//focas/focas.ps.Z}). This program offers a splitting routine which deals well with galaxy compounds.

With Source Extractor (SExtractor) we define the extension area of the galaxy. For this, through an iterative process, the background on the images was determined to identify objects with a signal higher or equal to three times this value. The  minimal area to be considered was imposed to be 3 $\times$ FWHM of the HST/ACS images to avoid the detection of spurious sources such as hot pixels and cosmic rays. 

At this step, different parameters are calculated for the detected objects. In particular, we measured the central position and the equivalent radius, i.e., the radius of a circle with the same area than that covered for the pixels associated with the object. Then the galaxy is relocated in the center of the image using the new position, and the image is resized to eight times its equivalent radius. The outermost isophote was used to obtain the flux, ellipticity and radius of galaxies which are provided in Table 3.

\subsection{Surface brightness analysis and photometry}

With the resized and centered images we performed an isophotal analysis with FOCAS. First, we calculated the background $\sigma$. Then, the objects are identified where the count level is above 3 times this value. We used the new position calculated for the objects to identify our target and separate it from other spurious sources that may be present in the field. For this, we define an area centered in the object and enclosed by an isophote with signal above 3$\times \sigma$ (see Fig. \ref{mosaico_03}; left panel). As a result, for each catalogued source we obtain, coordinates of the central position, total flux, nearby background, surface brightness, equivalent radius, and ellipticity. In Table 3 we show the apparent and absolute magnitudes,  luminous radii, colors and their errors, luminosities in H$\alpha$ and [OIII] and their errors, equivalent widths in H$\alpha$ and [OIII], surface brightness, ellipticity, and mass for each galaxy.

\subsection{Resolved star-forming regions and diffuse emission}

From visual inspection of the galaxies, different star-forming knots embedded in the more diffuse and extended emission are clearly visible in the galaxy images. Further analysis was carried out with FOCAS, with the aim of parametrizing such structures. For this, we find the signal over the diffuse emission of the galaxies, looking for individual regions. Besides, some regions may overlap, and then a criterion has to be defined to separate them.  The splitting of merged regions is done using two main parameters: a minimal area according to the FWHM of the PSF and a signal threshold. The signal threshold is set to be 3 $\sigma$ of the local background, which is computed iteratively from pixels that are not part of the object (see Valdes 1982, Sect. 4). A loop is entered incrementing the detection threshold level by 0.2$\sigma$ every time. Then, if at some iteration of the loop there is more than one region, the analysis continues separating each one. The loop finishes when the peak of every region is reached.

\begin{figure}[htbp]
\centering
\includegraphics[width=9.0cm]{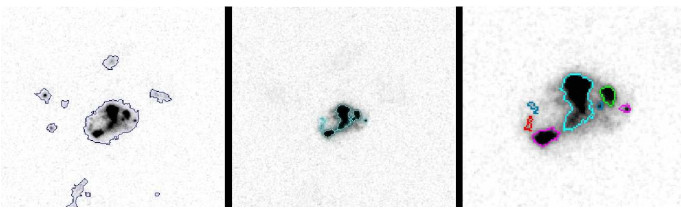}
\caption{{\small  F814W HST/ACS images. Example of the procedure to identify SF knots. Left: First isophotal analysis, where the principal objects in the image are defined. Middle: Second isophotal analysis of the target galaxy, where substructures are found. Right: A zoom of the image in the middle.For clarity the detected regions are showed with different colors.}}
\label{mosaico_03}
\end{figure}

In Fig. \ref{mosaico_03} the  process to identify substructures is shown. The left panel is the resized image with the objects detected (first isophotal analysis) in the field, with our target in the center. The  middle panel shows the image with external sources in the field removed. Subregions detected within the target are overplotted. The right panel is a zoom of the middle image. Regions with a radius bigger or equal to the PSF (0.09") will be identified as subregions or knots.

\begin{figure}[htbp]
\centering
\includegraphics[width=9.0cm]{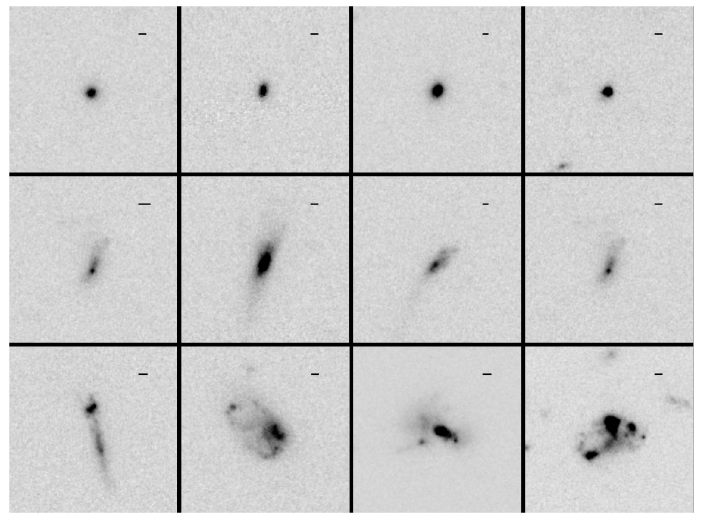}
\caption{{\small Example of the starburst morphological classes in our sample, a 1 kpc bar is showed at the top right of each image. Sknot galaxies (top row), Sknot+diffuse galaxies (middle row), and Mknots galaxies (bottom row).}}
\label{mosaico_classes}
\end{figure}

All starburst galaxies in our sample were morphologically classified based on this isophotal analysis as follows: Sknot, when it consists of a single knot of star formation; Mknots, when several knots of star formation are identified within the galaxy size; Sknot+diffuse, if the single knot is surrounded by diffuse emission. Fig. \ref{mosaico_classes} shows examples of the three morphological classes.

\begin{figure}[htbp]
\centering
\includegraphics[width=9.0cm]{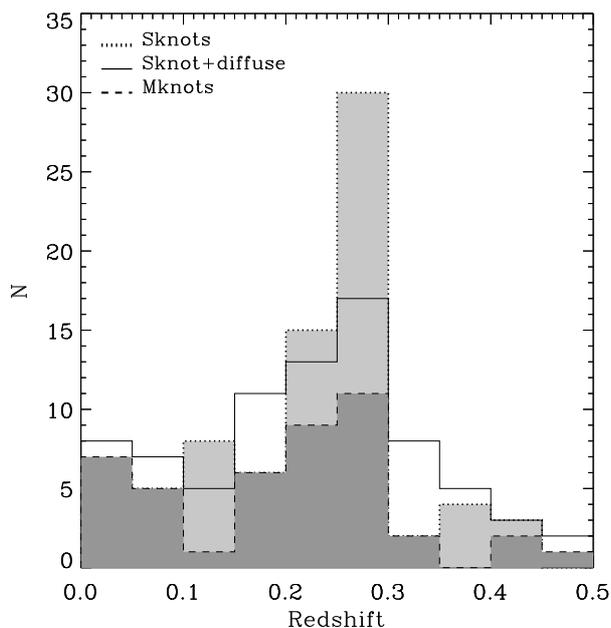}
\caption{{\small  Redshift distribution for the three starburst morphological classes defined in this paper: Sknot galaxies (light grey), Sknot+diffuse galaxies (empty histogram), and Mknots galaxies (dark grey).}}
\label{hist_redshift}
\end{figure}

The F814W filter was used for the morphological classification. This assure the H$\alpha$ emission is covered by the filter for all galaxies in our sample with $z>$0.1. At lower redshift, the clumpy morphological features will be detected by the continuum excess associated of the star-forming regions. The redshift distribution of the different starburst classes is presented in Fig. \ref{hist_redshift}. They display very similar trends and they are also in good agreement with the redshift distribution for the whole sample (see Fig. \ref{hist_red_per}). Therefore, we consider the filter bandpass is not severely limiting our study.

\begin{figure}[htbp]
\centering
\includegraphics[width=9cm]{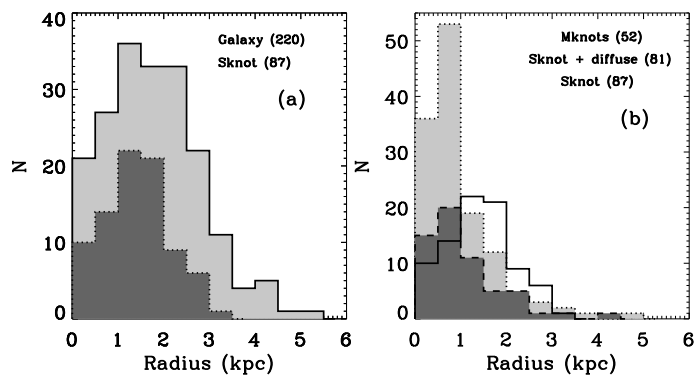}
\vspace{0.2cm}
\caption{{\small  Size distribution of the galaxies and star-forming knots from the isophotal analysis of the HST high spatial resolution F814W images. Circular symmetry is assumed to estimate the radius. Panel (a), the distribution of both the total sample and Sknot galaxies is shown in light grey and grey, respectively. Panel (b), the distribution of knots in Mknot galaxies, Sknot+diffuse galaxies, and Sknot galaxies is shown in light grey, grey, and empty histogram.}}
\label{tamanos_knots}
\end{figure}

We measured different parameters for each substructure: area, flux, magnitude, luminous radius and ellipticity. The luminous radius was determined assuming circular symmetry for the isophotes; this is a good approximation for the core of compact star-forming knots. Fig. \ref{tamanos_knots} shows the size distribution of galaxies and knots. The typical radius of the galaxies (Fig. \ref{tamanos_knots}a) and the knots of Sknots (Fig. \ref{tamanos_knots}b), Sknot+diffuse (Fig. \ref{tamanos_knots}c) and Mknots (Fig. \ref{tamanos_knots}d)  are 1-3 kpc, 1-2 kpc, 0.5-1.5 kpc, and 0.5-1.0 kpc, respectively.

In Table 4 we summarize the parameters of the knots in the starburst galaxies. As explained above, they are catalogued as Sknot, Sknot+diffuse and Mknots. When more than one knot in a galaxy was found, then they appear in the table following their H$\alpha$ luminosity from higher to lower values. The diffuse luminosity of each clump is obtained from the diffuse light of the galaxy, considering the area of the knot. The mass is calculated using $B-R$ color, using the prescriptions of \citet{bell2001}, assuming the same color for the knots and the host galaxy. Distance to the center, radius and ellipticity were determined with the isophotal analysis.

\begin{figure}[htbp]
\centering
\includegraphics[width=9.0cm]{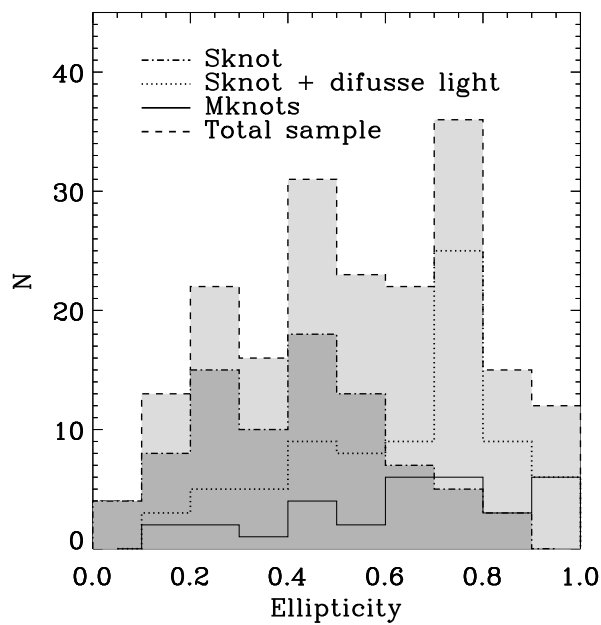}
\caption{Distribution of the ellipticity of the galaxies. Four cases are considered: galaxies with Sknot, Sknot+diffuse, Mknots and the total sample.}
\label{elip}
\end{figure}

In Fig. \ref{elip} we show the ellipticity of the galaxies for the different morphologies: Sknot, Mknots, and Sknot+diffuse. It is worth  noting that Sknot galaxies are preferentially round, with a mean ellipticity around 0.4. Galaxies with a single knot and diffuse light, and those with multiple knots, are more elongated, with mean ellipticities of 0.63 and 0.65, respectively.

\subsection{Spatial distribution of the star-forming knots in the galaxies}

\begin{figure}[htbp]
\centering
\includegraphics[width=0.49\textwidth]{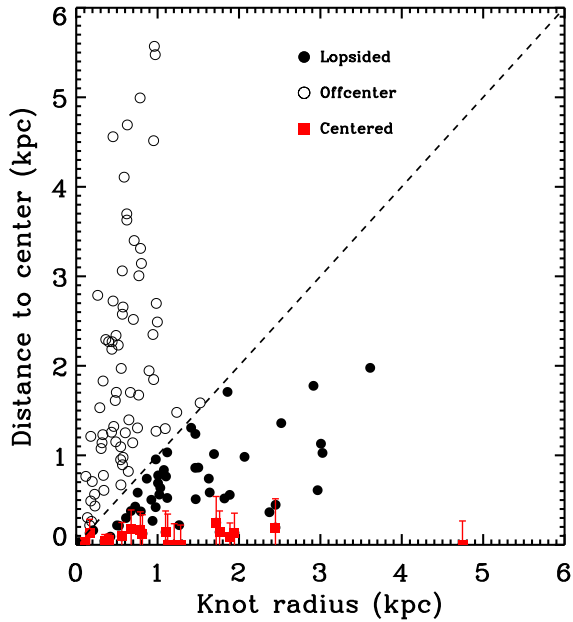}
\caption{Knot sizes versus their distance to the center of the galaxy. From the diagram we separate three classes: offcenter (open circles), lopsided (filled circles) and centered (red squares with error bars) regions. A bisector separates two regions. Filled circles show regions in contact with the center of the galaxy and open circles show regions offcentered. Regions which overlap with the geometrical center of the galaxy are labeled "centered" and represented with squared symbols. Overplot bars account for the spatial resolution. }
\label{radio_distancia}
\end{figure}

The galaxy centers were calculated with SExtractor, using the barycenter of the emission within the outer isophote; the center of the knots correspond to the maximum luminosity of the isophote, and their radii are equivalent to a circular shape of the isophotal area. All the projected linear scales have a resolution of 0.09" (limited by the PSF). Fig. \ref{radio_distancia} shows the distance to the center of the galaxy for each knot versus its radius. A bisector separates the knots in two classes: knots in the upper region (open circles) are offcentered, whereas in the lower region (filled circles) the knot overlaps with the geometrical center of the galaxy. We call them offcenter and lopsided, respectively. It is worth noting that those knots whose distance to the center is lower than their size - taking into account the spatial resolution of the HST - have been labelled as centered and plotted with a square symbol.

\begin{figure}[htbp]
\centering
\includegraphics[width=9.0cm]{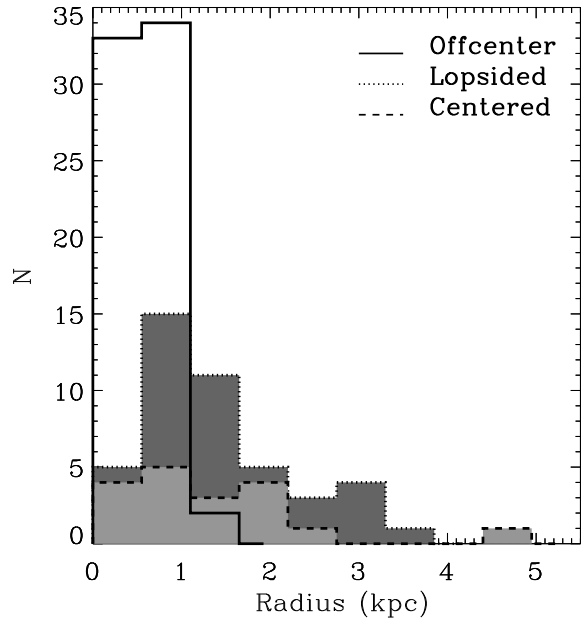}
\caption{Size distribution of knots, as identified in Section 4.3. Solid (empty), dotted (dark grey) and dashed (light grey) lines show the size distribution of offcenter, lopsided, and centered knots. The largest star-forming knots are located in the central regions of the galaxy.}
\label{size_distribution}
\end{figure}

Fig. \ref{size_distribution} shows the size distribution of the knots. Solid, dotted and dashed lines show the distribution of offcenter, lopsided and centered knots. Offcenter knots are smaller and more abundant. Lopsided and centered knots are generally larger but they also span a wider range in sizes. The mean radii are 0.1, 0.5, and 2.1 kpc for offcenter, lopsided, and centered knots, and the mean distances to the center are 1.3, 1.4 and 0.6 kpc, respectively. 

\subsection{Mass and SFR estimation of the star-forming knots}
\label{mass_sfr}

Individual star-forming knots/clumps in our galaxy sample cannot be resolved using the broad-band imaging provided by the SUBARU telescope. Therefore, in order to estimate their masses using the prescriptions given by \citet[][see Sect. \ref{sec:mass}]{bell2001}, we need to rely on the HST imaging. Unfortunately, the COSMOS field is only covered in its entirety by the F814W filter, so a clump-scale color is not directly accessible from the observations. In order to overcome this problem we assumed that the knots and the host galaxy have the same color. Then, we used the F814W-band images to measure the luminosity of the individual knots and the SUBARU color to derive their corresponding masses. 

To test the influence of our color hypothesis in the derived knot masses we cross-matched our sample with CANDELS \citep{grogin2011,koekemoer2011} and 3D-HST \citep{brammer2012} surveys. These new HST surveys only cover a small area of the full COSMOS footprint but they provide imaging using different filters at cluster-scale resolution. We found 4 starburst galaxies classified as Sknot+diffuse or Mknots galaxies in this search, accounting for a total of 12 knots/clumps of star formation. Using the F606W filter provided by CANDELS we computed the colors of the individual knots as well as those of the entire galaxy. The mean value of the rest-frame F606W-F814W (equivalent to $V-I$) color is 0.1 for both the knots and galaxies. However, the knots present a broader range of values ($\sigma\sim$ 0.5 mag). Despite the small sample used in this study (only 4 galaxies and 12 knots), our results are backed up by previous findings in the literature. \citet{wuyts2012}, using a sample of 323 starburst galaxies, claimed that no difference in color is found between the disk and clump regions when the clumps are detected using the surface stellar mass as detection method (see their Fig. 6). Guo et al. 2012 found also a similar result using a sample of 40 clumps detected using $z-$band images. They claimed that the mean $U-V$ color of clumps is similar to that of disks, with the color distribution of clumps being broader (see their Fig. 5). Therefore, even if for a given knot the color difference can be up to 0.5 magnitudes (1$\sigma$), on average the assumption that the colors of knots and the host galaxies are the same is still valid. To account for the individual variations we have propagated a 1$\sigma$ error in the clump colors to the derived stellar masses. Still,  possible  radial  color   gradients  can  be  present  within individual galaxies.  \citet{guo2012} estimated a $U-V$ color difference between  center   and  offcenter   star-forming  knots   of  $\sim$0.5 magnitudes,  similar   to  the  1   $\sigma$  scatter  for   our  mean colors. Whether this color gradient is present in our sample, it would  systematically overestimate  the stellar  mass for  bluer knots (bluer than  the integrated  galaxy) and  underestimate it  for redder knots by a factor comparable to our 1 $\sigma$ error.

Another possible caveat on our knot mass determination might be the use of different band-passes affecting our results. In order to estimate how accurate is this approximation we use our sample of Sknot galaxies. These starburst galaxies do not contain diffuse light and, therefore, the measurement using the higher resolution HST images should agree with those from the ground-based SUBARU photometry. We calculated the mass of the galaxies using the $M/L$ ratio \citep{bell2001} with the $B-R$ color (the less affected by the background galaxy), and the luminosity in the SUBARU $I-$band and HST/ACS F814W-band with a fixed aperture of 3" (COSMOS catalogue). We find an excellent agreement between both measurements.

\begin{figure}[htbp]
\centering
\includegraphics[width=9.0cm]{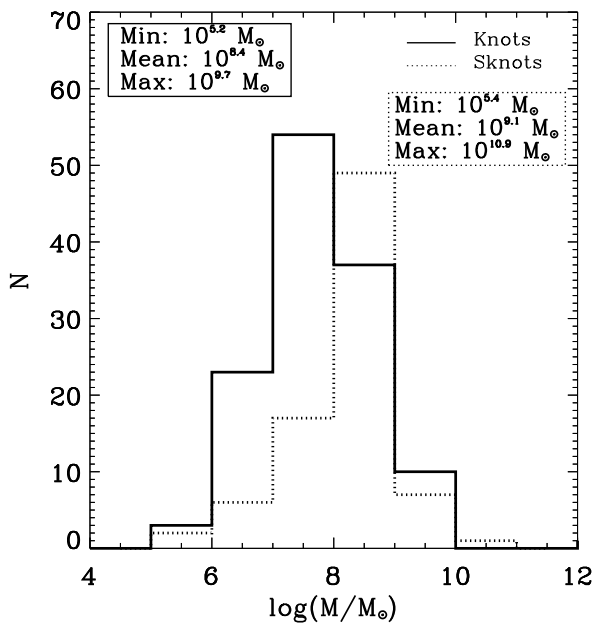}
\caption{Mass distribution of the knots in Sknot+diffuse, and Mknots galaxies (solid line) and Sknot galaxies (dashed line). The statistics is given in the boxes with the same line code.}
\label{mass_knots}
\end{figure}

\begin{figure}[htbp]
\centering
\includegraphics[width=9.0cm]{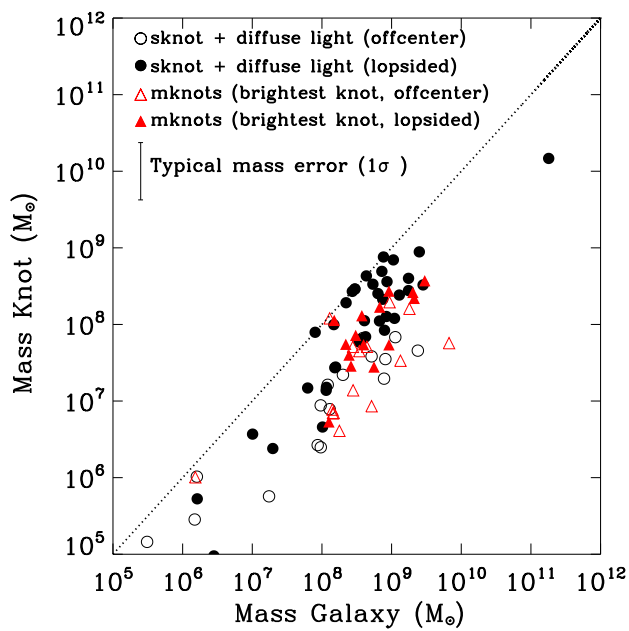}
\caption{Comparison of masses of the knots with that of their host galaxies. Symbols represent the different morphologies (code inserted): One knot plus diffuse light, and the brightest knot in multiple knot galaxies. Sknot galaxies are not represented, as they would follow the bisector.}
\label{comp_mass_dist}
\end{figure}

The masses of the knots and their corresponding errors are provided in Table 4 and represented in Fig. \ref{mass_knots}. The mean  mass for knots in Sknot+diffuse and Mknots galaxies  in 10$^{8.4}$ M$_\sun$. The minimum and maximum are 10$^{5.1}$ and 10$^{9.7}$ M$_\sun$, respectively. We also show the mass distribution of the single knot galaxies (dashed line). We compute a mean mass for Sknots of 10$^{9.1}$ M$_\sun$, 5 times larger than the mean of knots in Sknot+diffuse and Mknots galaxies. For Sknot+diffuse and Mknots, we compare the masses and sizes of the knots and the galaxies. The mass comparison is  shown in Fig.~\ref{comp_mass_dist}. The masses  of the knots in Sknot and Mknots galaxies correlates  with the total galaxy mass, with more  massive knots  being  in more  massive  galaxies. As  previously pointed out,  a systematic  color gradient in  our galaxy  can produce differences in  the mass estimation of  the knots in Mknot  galaxies of the order of the typical mass error shown in Fig.~\ref{comp_mass_dist}.

The H$\alpha$ luminosity of the different star-forming galaxies in our sample was estimated following different strategies. Sknot do not require the HST/ACS spatial resolution to be measured. Therefore, we used the SUBARU intermediate filters to determine the luminosity in H$\alpha$ and continuum regions. Using these filters we were able to trace the H$\alpha$ emission in the redshift ranges 0.01 $\le z \le$ 0.2 and 0.23 $\le z \le$ 0.28. 20 Sknot galaxies in our sample are in this redshift range, and nine of them have spectra in zCOSMOS. We used the difference between photometric and spectroscopic H$\alpha$ luminosity of these galaxies to estimate our uncertainty, the mean of this difference is 0.66 dex.

For Sknot+diffuse and Mknots with $z>0.1$ the F814W filter includes the H$\alpha$ emission, the continuum associated to the star-forming region, and the continuum+H$\alpha$ emission of the diffuse light of the host galaxy. The continuum and H$\alpha$ diffuse emission from the host galaxy was removed computing the luminosity in an area equivalent to the knot but in the diffuse gas region of the galaxy and subtracting it to that of the corresponding knot. 

\begin{figure}[htbp]
\centering
\includegraphics[width=8.0cm]{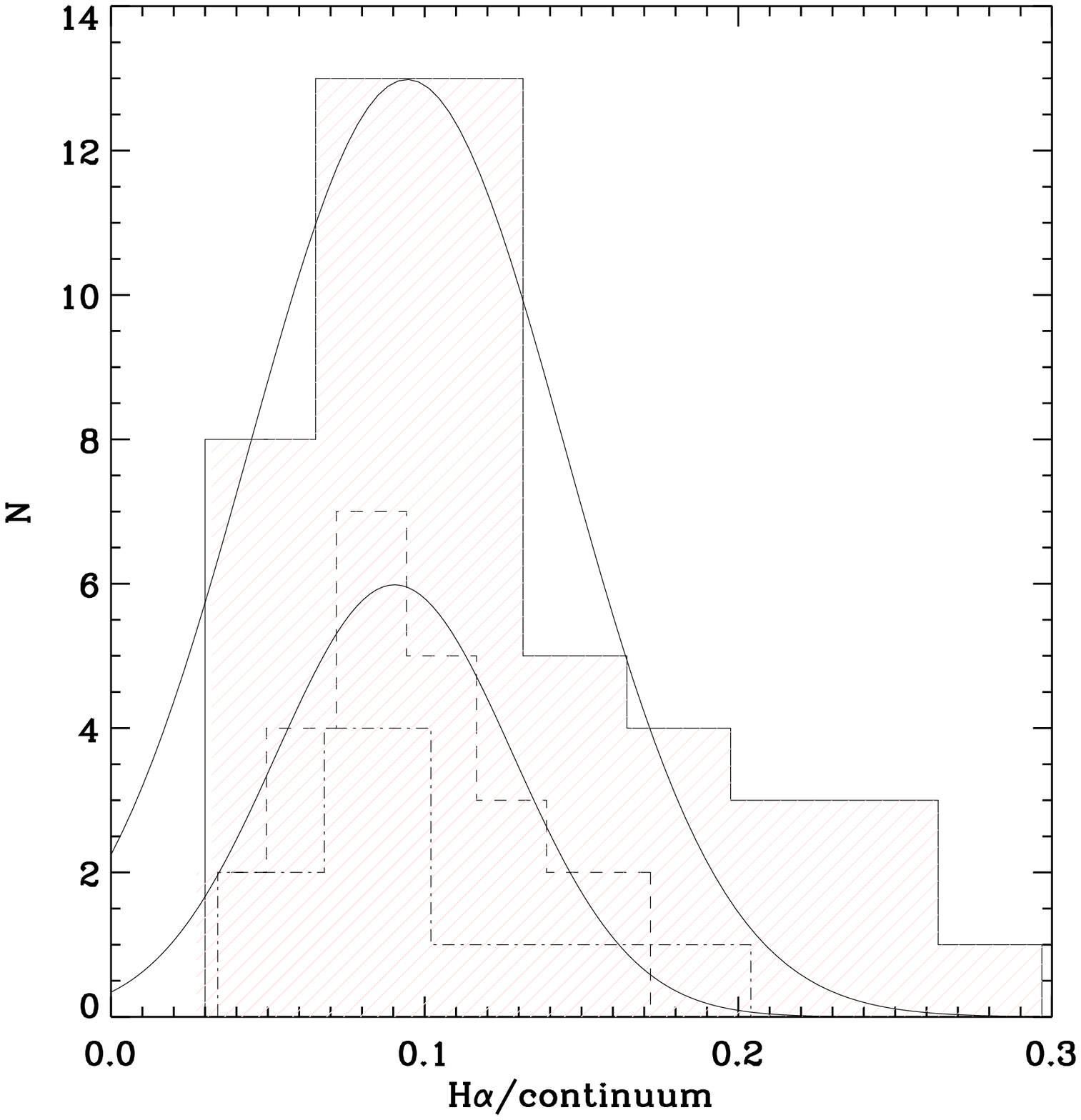}
\caption{H$\alpha$/continuum distribution for galaxies in our sample with spectra in zCOSMOS. Different lines show: the distribution of the 51 Sknot+diffuse and Mknots galaxies (solid line) with a fitted Gaussian profile overplot; the distribution of a subsample of 9 galaxies with bright knots (dot--dashed line)- its luminosity twice or more that of the host galaxy, and the distribution of the 25 Sknot galaxies (dashed line) with the fit with a Gaussian overplot.}
\label{lum_knots}
\end{figure}

The star-forming continuum was subtracted using an statistical value derived from the  H$\alpha$/continuum ratio for 51 Mknots galaxies with spectrum in zCOSMOS. Fig. \ref{lum_knots}  shows the distribution of this 'correction' ratio. The Gaussian fit provides a mean of 0.09, with a dispersion of 0.05. We used the mean value to correct the H$\alpha$ luminosity of the knots and the dispersion was included in the error budget of the derived quantities (in particular of the SFR). A possible caveat to this method is whether H$\alpha$+continuum of the host galaxy could be present in zCOSMOS spectra. To check this issue, we used a subsample of galaxies with available spectra and with a star-forming knot/clump luminosity at least twice the luminosity of the host galaxy (9 galaxies). For this subsample we would expect to have a negligible contribution from the host galaxy due the different luminosities. We found remarkably similar results using this small sample (see Fig. \ref{lum_knots}, dot-dashed line). We also analyzed a subsample of 25 Sknot galaxies (galaxies without diffuse light) and with available spectroscopy in zCOSMOS. For this sub-sample we would expect our H$\alpha$/continuum distribution to be shifted towards lower values since there is no contribution from the host galaxy. However, we find again a good agreement with our initial sample of 51 galaxies. Fig. \ref{lum_knots} (dashed line) show the derived distribution for Sknot galaxies. The best Gaussian fit to this distribution provides also a mean of 0.09 and a dispersion of 0.04.

Therefore, we consider that a H$\alpha$/continuum correction factor of 0.1 with a typical dispersion of 0.05 is a robust value even in the most extreme cases in our sample. The final luminosity in H$\alpha$, obtained as described above for Sknot, Sknot+diffuse and Mknots galaxies, is given in Table 4. 

\begin{figure*}[tH]
\centering
\includegraphics[width=14.0cm]{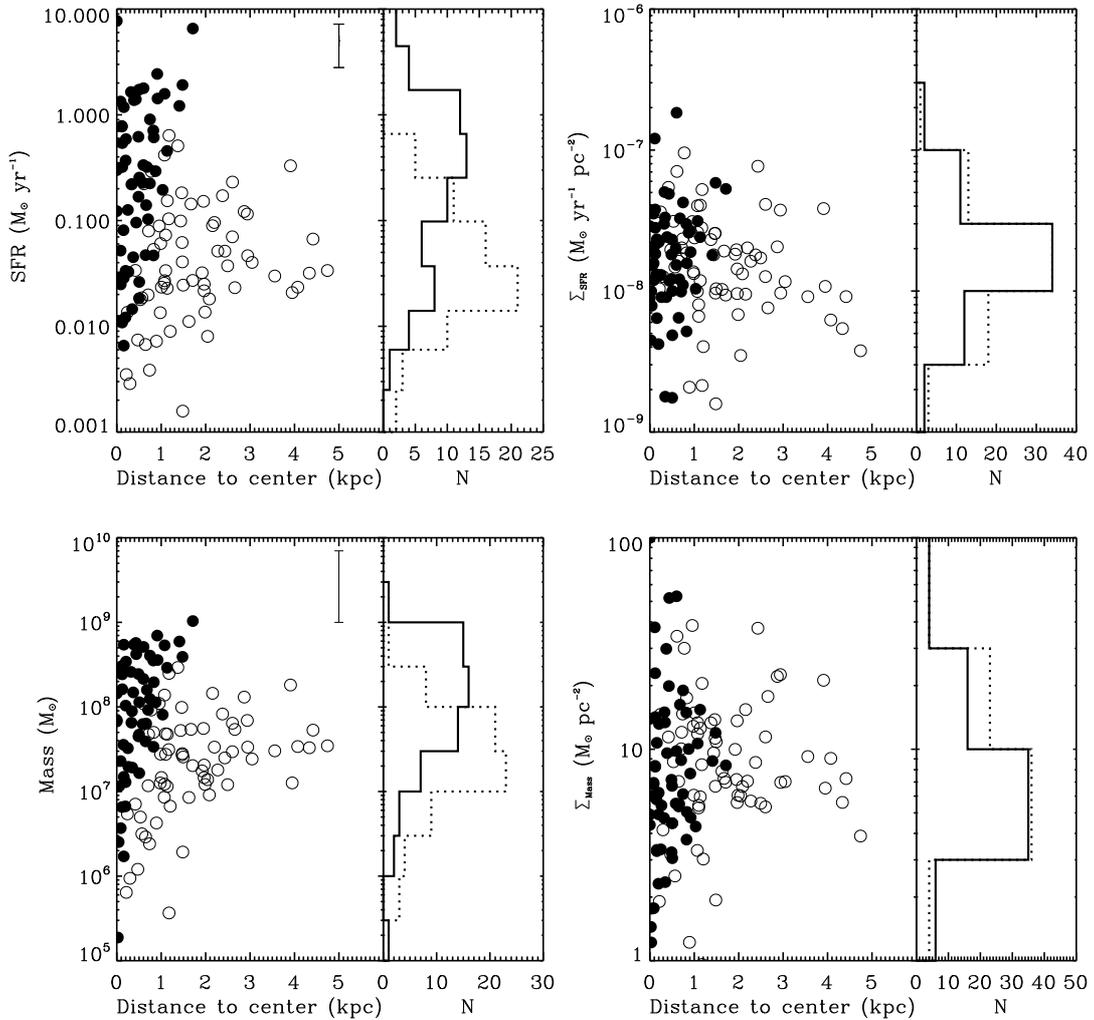}
\caption{SFR and $\Sigma_{SFR}$ versus galactocentric distance (top), and mass and $\Sigma_{mass}$ versus galactocentric distance (bottom). Filled circles are centered knots and open circles represent offcenter knots. The distribution of each quantity in the $y-$axis is showed at the right of each panel, solid line is for centered knots and dotted line is for offcenter knots. As it can be seen, the most massive and high star-forming regions are in the central part of the host galaxies. For the sake of clarity Sknots have not been display.}
\label{ssfr_distance}
\end{figure*}

The SFR is calculated from the luminosity in H$\alpha$ using the relation between L(H$\alpha$) and SFR from \citet{kennicutt1998}. Fig. \ref{ssfr_distance} shows  the radial distribution of the SFR, SFR/area, mass and mass/area for centered, offcentered, and lopsided knots, as defined in Section 4.3. As can be seen in Fig. \ref{ssfr_distance} (top) the knots with the highest SFR ($\sim$1 $M_\sun$ yr$^{-1}$) are only present in the central regions of the galaxies. Fig. \ref{ssfr_distance} (bottom) shows the mass and mass/area versus the distance to the center. The more massive and largest star-forming regions are also in the central part of the host galaxies. It  is worth  noting that  whether  internal color  gradients of  the star-forming knots  would be present  in our sample,   the statistical mass difference between both  populations (center and offcenter knots) would be larger.

\subsection{Single Knot Galaxies}

\begin{figure}[htbp]
\centering
\includegraphics[width=9.0cm]{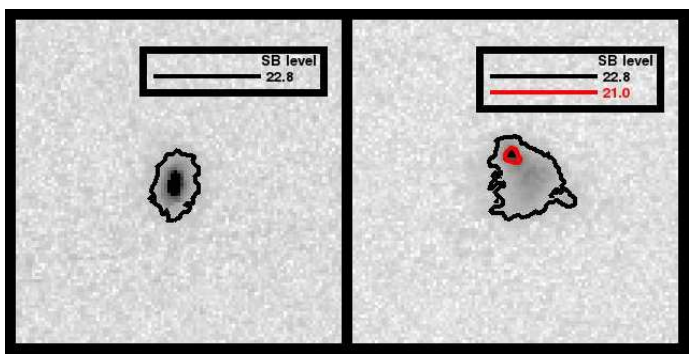}
\caption{Substructures in a Sknot (left panel) and Sknot+diffuse (right panel) galaxy. Isophotes are in magnitude/arcsec$^2$. The analysis have not found substructures and diffuse light in the Sknot galaxy, while for Sknot+diffuse galaxy it is clearly seen how substructure and diffuse light is found far from the knot region.}
\label{mosaico_sknots}
\end{figure}

In our catalogue, 87 galaxies were classified as Sknot ($\sim$38\%), meaning that they show a resolved star-forming region and no extended diffuse emission. The minimal area used to define a substructure is 15 connected pixels, with values over 3$\sigma$ the sky background. This gives us a magnitude limit of $\sim$26. This faint magnitude would allow us to detect very low surface-brightness sources. For this reason, and within this magnitude limit, we are confident that Sknot galaxies have no diffuse extended light. Fig. \ref{mosaico_sknots} shows two examples of the search for substructures in Sknot (left panel) and Sknot+diffuse (right panel) galaxies. The units are in magnitude/arcsec$^2$. 

Sknot galaxies are larger (see Fig. \ref{tamanos_knots}) and more massive (see Fig. \ref{mass_knots}) than star-forming knots in Sknot+diffuse and Mknots galaxies. However, they have similar $\Sigma_{SFR}$ and $\Sigma_{mass}$, implying that they probably share a similar SF mechanism. 

The ellipticity distribution of Sknot galaxies shows that they are rounder than other starburst categories, (see Fig. \ref{elip}). The mean is 0.4, which is in agreement with spheroidal galaxies viewed from random line-of-sights \citep{padilla2008, mendezabreu2016}.

\subsection{SFR-stellar mass relation}

Figure.~\ref{mass_sfr} shows  the SFR-stellar  relation for  our sample  galaxies,  as  well   as  that  of  individual  star-forming
  knots. SFR  and stellar masses  measurements have been  described in Sects.~\ref{sec:mass} and  \ref{sec:morph}. The  best linear  fit to the whole galaxy  sample and the subsample of  star-forming knots is shown  in  Fig.~\ref{mass_sfr}.  The slope of the fit for our targets (both the galaxies and the knots) is 0.68.  We have compared this relation to that previously presented by \citep{whitaker2012} for star-forming galaxies at z=0. The slope of their fit at this redshift is 0.7. We found that our SFR-stellar mass  relation for the  global galaxy properties  lie in the locii of other star-forming  galaxies studied in the literature. We also  found that even if the clumps follow the  same SFR-stellar slope as the galaxies,  their SFR is systematically larger by a  factor  of  $\sim  3$. This result  is  compatible  with  that presented by \citep{guo2012}  where they measure the SFR in the knots that resulted to be a factor of 5 higher than that of the underlying disks. Note that the SFR that we measure for the galaxies includes also that corresponding to the starburst knots, what can explain that the difference in the SFR (clumps-galaxies) in our data is smaller.

\begin{figure}[!h]
\centering
\includegraphics[width=9.0cm]{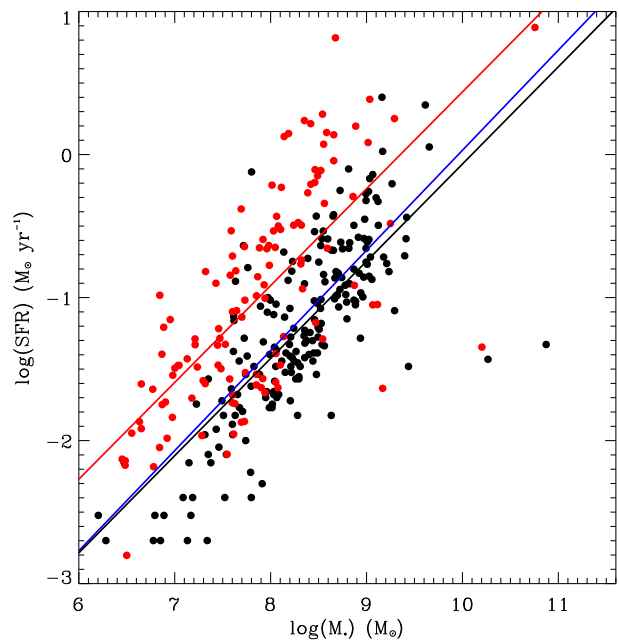}
\caption{SFR–stellar  mass relation  for  our  sample galaxies  (black dots)  and star-formation  knots  (red dots).  The SFR–stellar  mass  relation of each population is fitted by a straight line (black and red). The blue solid line represents the relation of \citet{whitaker2012} for star-forming galaxies at $z$=0.}
\label{mass_sfr}
\end{figure}

\section{Discussion}
\label{sec:discussion}

\subsection{Comparison with the literature at this redshift}
\label{comp_lit}

The number and properties of star-forming knots in a starburst galaxy provides important information to test the different cosmological models predicting the formation and evolution of galaxies. The availability of recent deep surveys has lead to several works aiming at quantifying the regions in the galaxies were SF is occurring. The precise definition of a starburst knot, however, is very much dependent on the particular observations. An unique definition to search for starbursts in deep fields is still lacking, and the parameters used so far have varied in the different works published in the literature. First studies were done visually \citep[e.g.,][]{cowie1995, vandenbergh1996, abraham1996, elmegreen2004a, elmegreen2004}.  More recently, \citet{guo2014} used the excess in UV as the physical parameter to identify a clump. In particular, they classified as a star-forming clump those contributing more than 8$\%$ to the rest-frame UV light of the whole galaxy. Using this definition, they measured the knots in star-forming galaxies in the redshift range 1 < $z$ < 3. At lower redshift, 0.2 < $z$ < 1.0, \citet{Murata2014} identified the star-forming knots using the HST/ACS F814W-band in the galaxies in the COSMOS survey. The criteria used was that the detected sources must be brighter than 2$\sigma$ the local background, and with a minimum of 15 connected pixels. In this paper, we have also used the HST/ACS F814W-band in the COSMOS survey to find and parametrize the star-forming knots of the starbursts identified in the COSMOS survey. In particular, we calculated the emission ($\sigma$) of the host galaxies and searched for regions with 3 times this value, which is more strict than the 2$\sigma$ used by \citet{Murata2014}, over 15 connected pixels using FOCAS.

The clumpy fraction ($f_{clumpy}$; clumpy galaxies/SF galaxies) have been investigated for different redshift and mass ranges \citep{Elmegreen2007, overzier2009, puech2010, guo2012, wuyts2012, guo2014, Murata2014, tadaki2014}. Despite the already mentioned differences in the clump definition, a remarkable agreement has been reached, concluding that $f_{clumpy}$ increases with redshift. However, a direct comparison between different works is often difficult, not only because of the clump definition, but also due to the different definition of a starburst galaxy. For example, \citet{guo2014} and \citet{Murata2014} defined their samples selecting galaxies with specific SFR (sSFR) $\ge$ 0.1 Gyr$^{-1}$.  In other works, such as \citet{guo2012}, they used sSFR $\ge$ 0.01 Gyr$^{-1}$. \citet{wuyts2012} used galaxies that need less than the Hubble time to form their masses. \citet{tadaki2014} selected galaxies with H$\alpha$ excess and  \citet{overzier2009} used Lyman Break Analog (LBA) galaxies, which share typical characteristics of high-redshift Lyman Break Galaxies (LBG). On the other hand, \citet{puech2010} used [OII] emission-line galaxies. The samples are then defined using a variety of criteria. In this paper we used the values of EW(H$\alpha$) and EW([OIII]) $\ge$ 80 \AA\ to search for star-forming galaxies. These parameters are used in nearby HII regions and starburst galaxies, and points directly to recent (young) star formation \citep{kniasev2004,cairos2007,cairos2009a,cairos2009b,cairos2010,morales2011,amorin2014}.

\begin{figure*}
\centering
\includegraphics[width=\textwidth]{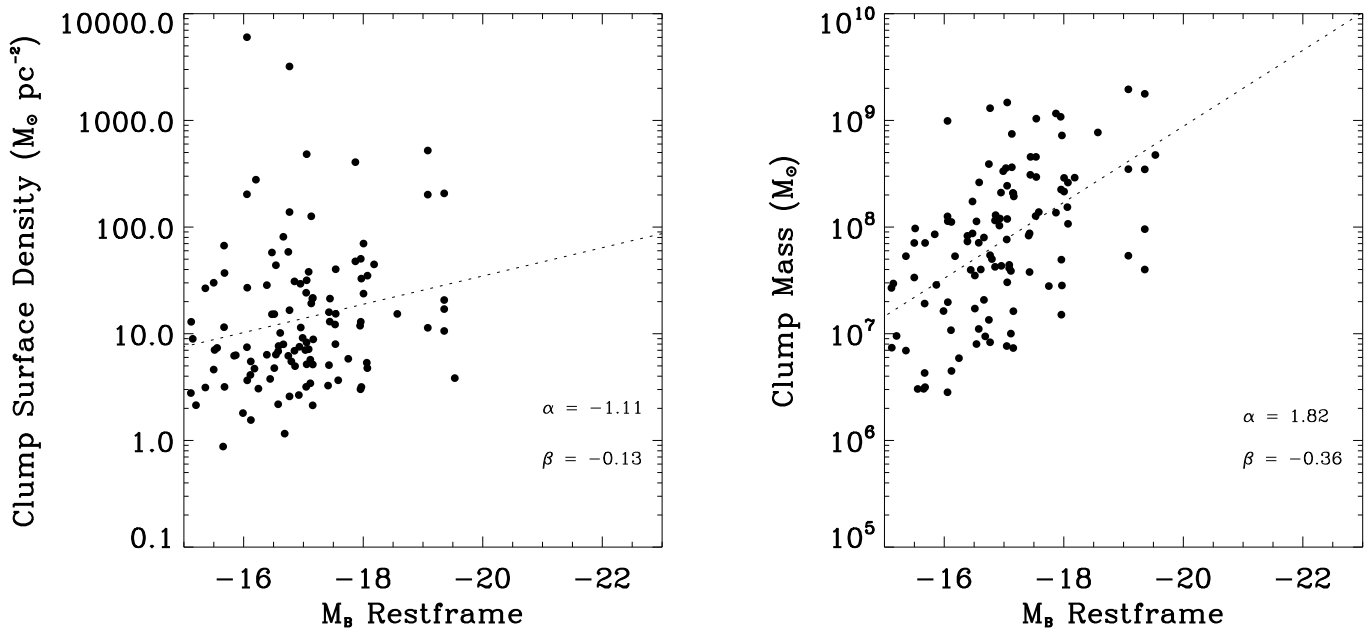}
\caption{Clump Surface Density (left) and Clump Mass (right) versus absolute $B-$band magnitude of their host galaxy. The best fit to the data is shown in dotted lines. The values of slopes and intercepts are giving in the lower right corner.}
\label{comp_mass}
\end{figure*}

Using our definition of both starburst galaxy based on the EW of the H$\alpha$ and [OIII] lines and star-forming clump based on light excess in the F814W filter, we calculated the clumpy fraction for our galaxies. Accounting for our entire sample up to $z\sim0.5$ we found a value of $f_{clumpy}$=0.24. The mass range of the galaxies in our sample is 10$^6$-10$^9$ M$_\odot$ (Sect. \ref{sec:mass}), while in previous studies these values are usually larger \citep[10$^9$-10$^{10}$ M$_\odot$]{overzier2009}, \citep[>2$\cdot$10$^{10}$ M$_\odot$]{puech2010}, \citep[>10$^{10}$ M$_\odot$]{guo2012}, \citep[>10$^{10}$ M$_\odot$]{wuyts2012}, \citep[10$^9$-10$^{11.5}$ M$_\odot$]{guo2014}, \citep[10$^9$-10$^{11.5}$ M$_\odot$]{tadaki2014}, \citep[>10$^{9.5}$ M$_\odot$]{Murata2014}. The value obtained in this work for $f_{clumpy}$ is larger than the value 0.08 found by \citet{Murata2014} at $z$=0.3. Besides the differences in the definition of the knots, the lack of agreement in the value of $f_{clumpy}$ can be explained by the difference in the mass range of the galaxy sample. On the other hand, \citet{Murata2014} defined as clumpy a galaxy with 3 or more knots, while we have considered galaxies with 2 or more knots. It is worth noting that other caveats discussed throughout this paper, such as bandpass or spatial resolution effects, would only increase our clumpy fraction.

\citet{Elmegreen2013} compared the properties of individual clumps of three different galaxy samples: knots in local spiral galaxies (obtained from the Sloan Digital Sky Survey; SDSS) massive clumps in local starbursts belonging to the Kiso Survey \citep{miyauchi2010}, and clumps in galaxies at high redshift from the Hubble Ultra Deep Field  (HUDF). They found correlations between different parameters of the clumps and the absolute $B-$band magnitude of the host galaxy.  For a comparison with \citet{Elmegreen2013}, we used two properties of our sample knots: surface density and mass. In Fig. \ref{comp_mass} we plot them as a function of  absolute $B-$band magnitude of the galaxy. These plots are similar to those shown in Figs. 4 and 6 in \citet{Elmegreen2013}. The best fit was determined for each parameter, and in the lower right corner of the figures the values of the intercept and slope are given. The surface density and mass trends are in between those provided in \citet{Elmegreen2013} for local and high-redshift massive clumps. The mass versus absolute $B-$band magnitude slopes for Kiso, HUDF and our sample are -0.54, -0.18 and -0.36, respectively. For surface density versus absolute $B-$band magnitude the values are -0.18, -0.12 and -0.13. The values suggest that for a given absolute $B-$band magnitude of the galaxy, the mass and surface density of the knots of the sample of this paper have higher values than those of clumps in local spirals, and lower than those found in high-redshift massive star-forming regions. There is a shift in the intercept of the linear regression fit of our data points, which has higher values than those provided in \citet{Elmegreen2013}. This might be due to the different definitions to determine the area to retrieve the mass, surface density and SFR.

\subsection{Scaling relations}

Scaling relations are a useful tool to determine parameters when the spectral or spatial resolution is a limitation. A particular case of interest in this work is when the spatial resolution is not good enough to resolve the star-forming regions. Then, it is usual to rely on the L(H$\alpha$) vs. diameter relation to estimate the size of the region we are interested in. In this reasoning, we are assuming that the same physical mechanisms and conditions can be applied to star-forming regions throughout a large range of luminosities and sizes.

In our sample, we took advantage of the HST high spatial resolution images, which allow us to spatially resolve the clumps of SF, providing an accurate estimation of their sizes. Furthermore, it allows us to estimate the L(H$\alpha$) for individual clumps and explore the L(H$\alpha$) vs. diameter relation without further hypotheses. We estimated the L(H$\alpha$) for clumps in galaxies with $z$ $\ge$ 0.1 as was explained in section 4.4. With these values we plotted the L(H$\alpha$) vs. diameter to explore this scaling relation. 

\begin{figure}[!t]
\centering
\includegraphics[width=9.0cm]{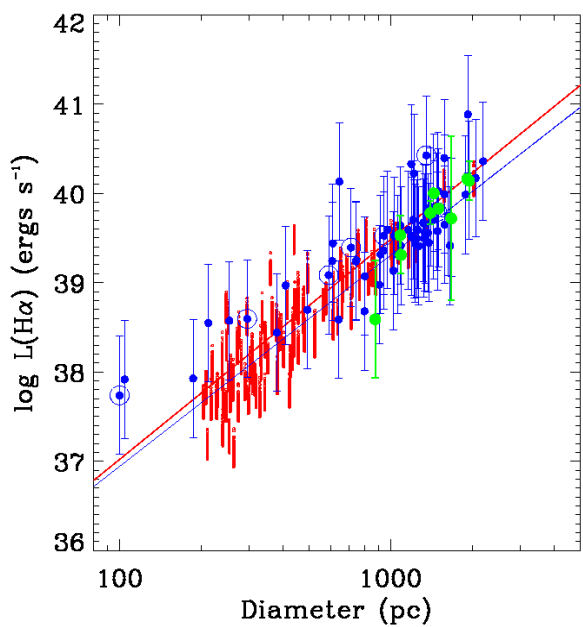}
\caption{L(H$\alpha$) versus diameter relation.  Blue points are Sknot galaxies. Circled enclosed points were discarded for the fit because of contamination in the flux (see text sec. 6.2). Green points are Sknot, with L(H$\alpha$) determined photometrically and spectroscopically; the error bar is the difference in luminosities. The mean value of the errors is taken as proxy error for the measurements (error bar for blue points). Red points are estimations of L(H$\alpha$) of clumps with a Monte Carlo simulation. The solid red line is the best fit for red points and solid blue line is the best fit for Sknot galaxies.}
\label{lum_rad}
\end{figure}

Fig. \ref{lum_rad}  shows the H$\alpha$ luminosity versus diameter of knots in our sample. Blue points are Sknot galaxies. Some Sknot galaxies do show field objects within the fixed SUBARU aperture ($\phi$=3"). These objects are shown enclosed by a circle, and have been discarded for the fit. Sknot with spectra in zCOSMOS were used to estimate errors associated with the luminosities as was explained in Sect. 4.4. Green points are these objects, and the error bar is the difference between the spectroscopic and photometric H$\alpha$ luminosity. The mean value of this difference is taken as the error for the blue points.

The continuum-corrected H$\alpha$ emission for Sknot+diffuse and Mknots is also represented in Fig. \ref{lum_rad} (red points). The errors associated to each knot were computed by propagating the uncertainty in the continuum correction as explained in Sect. 4.4. In order to estimate how the uncertainties associated with these H$\alpha$ measurements can influence the best-fit of the scaling relation, we ran a set of Monte Carlo experiments. We created 100 simulated distributions of H$\alpha$ luminosity of the knots. Each individual galaxy was allowed to vary its luminosity within the 1$\sigma$ correction obtained from Fig. \ref{lum_knots}. The best-fit for each distribution was obtained, and it is represented in Fig. \ref{lum_rad} (red line). The mean slope and dispersion obtained from this method was 2.46 and 0.04, respectively.

The value of the slope is an important parameter to understand the universality of the L(H$\alpha$) vs. diameter scaling relation. In our sample we found the relation log$(L_{H\alpha})=(32.2\pm0.1)+(2.48\pm0.05) \cdot$log(d) for Sknot, Sknot + diffuse and Mknots galaxies. \citet{Fuentes-Masip2000} obtained a value for the slope of 2.5 for giant HII regions in NGC4449, and \citet{Wisnioski2012} found a value of 2.78 using local giant HII regions and high-redshift clumps.

\subsection{Mknots galaxies and their link to bulge formation}

Several mechanisms have been proposed to explain the formation and evolution of galaxy bulges \citep{kormendy2004, bournaud2015}. At high redshift, bulge formation is thought to be driven by either gravitational collapse \citep{eggen1962}, major mergers \citep{hopkins2010}, or coalescence of giant clumps \citep{Bournaud2007}. This latter scenario, in which massive star-forming clumps migrate from the outer disk to the galaxy centre forming a bulge, has been extensively discussed using numerical simulations \citep{Noguchi1999, Immeli2004a, Immeli2004b, Bournaud2007, Elmegreen2008, ceverino2010}. Recently, \citet{Mandelker2014} used cosmological simulations to provide clump properties to be compared with observations. In particular, they separated clumps formed {\it in-situ} (clumps formed by disk fragmentation due to violent instabilities) and {\it ex-situ} (clumps accreted through minor mergers), finding differences in their properties and radial gradients that could help to distinguish their origins.

In  this scenario,  \citet{Mandelker2014} predict  that {\it  in-situ} clumps  show a  radial  gradient  in mass,  with  more massive  clumps
populating  the  central   regions.   The  aforementioned  theoretical studies were focussed  on massive galaxies. More  recent analysis deal with   less  massive   galaxies  by   means  of   similar  theoretical simulations. In \citet{ceverino2016} they follow gas inflow that feeds
galaxies with low metallicity gas from the cosmic web, sustaining star formation across  the Hubble  time. Interestingly, their  results show
clump  formation  in  galaxies  with   stellar  masses  of  $M  \simeq 10^9 $$M_{\odot}$  using zoom-in  AMR cosmological  simulations.  
Altough the cosmic  baryonic accretion rate  for  high-mass galaxies  significantly  drops  from $z\sim2$  to $z\sim0$  \citep[see][]{dekel2013}, the cold flow accretion into low mass galaxies does  not dramatically  change with redshift.  The  simulations in  Ceverino etal. (2016) with  cold gas inflows reaching the galaxy and feeding the SF in galaxies of masses $\sim 10^9$ $M_{\odot}$  are the  only
available so far. They can be considered a good example to compare with observational   information as the one provided in this paper. 

Fig. \ref{ssfr_distance} shows  the distribution of mass and mass  surface   density  for   our  clumps  as   a  function   of  the
galactocentric distance. Despite the errors  introduced by the lack of multi-band photometry at cluster-scale resolution, it is clear how the
most massive  clumps in our sample  are located in the  galaxy center, with offcenter  clumps being less  massive. However, this  tendency is not so clear in terms of  surface mass density, where some offcentered clumps have  similar surface  mass densities to  those located  in the center, thus likely implying an  {\it ex-situ} formation. On the other hand, numerical simulations  also predict that more  massive clumps of SF should appear in  more massive galaxies \citep{Elmegreen2008}. This is  nicely  reproduced  in  Fig.   \ref{comp_mass_dist},  and  it  is, therefore, consistent  with most of  the clumps being formed  at large radii, and then accreting mass from the disk as they migrate inwards.

Figure \ref{ssfr_distance} also shows the distribution of SFR and SFR surface density versus galactocentric distance. Centered clumps show slightly higher SFR than offcentered ones. This behavior would be expected in the case of {\it ex-situ} formation of the clumps \citep{Mandelker2014}. However, the fairly constant radial SFR surface density, also shown in Fig. \ref{ssfr_distance}, and predicted for both scenarios, prevented us from extracting further conclusions.

Figure.~\ref{mass_sfr} shows  the SFR-stellar  relation for the galaxies and  the individual  star-forming knots. The Figure also includes the  best linear fit to both samples and the one given by \citep{whitaker2012} for star-forming galaxies. The SFR-stellar mass  relation for the galaxies lie in the locii of other star-forming  galaxies studied in the literature. The clumps follow the  same SFR-stellar trend with the same slope but shifted systematically by a  factor of $\sim  3$.   This result  is consistent with  \citep{guo2012}  that reported a SFR in the knots to be a factor of 5 higher than that of their underlying disks.

In our sample we calculated the sSFR for each knot for every Sknot+diffuse and Mknots galaxy. Fig \ref{surface} shows the distribution of sSFR for knots in Sknot+diffuse and Mknots galaxies in our sample, with a bin width of 250 Myr$^{-1}$. We have also explored the sSFR for clumps which are centered and offcentered in Sknot+difffuse and Mknots galaxies. The values are similar, and no different SF mechanism can be identified based in the sSFR. Sknots and knots in Sknot+diffuse and Mknots galaxies have similar values of sSFR; the similitude in the distribution for both samples suggests the same mechanism of SF.

\subsection{Two populations: the knots that are galaxies and the remaining}

\begin{figure}[htbp]
\centering
\includegraphics[width=8.0cm]{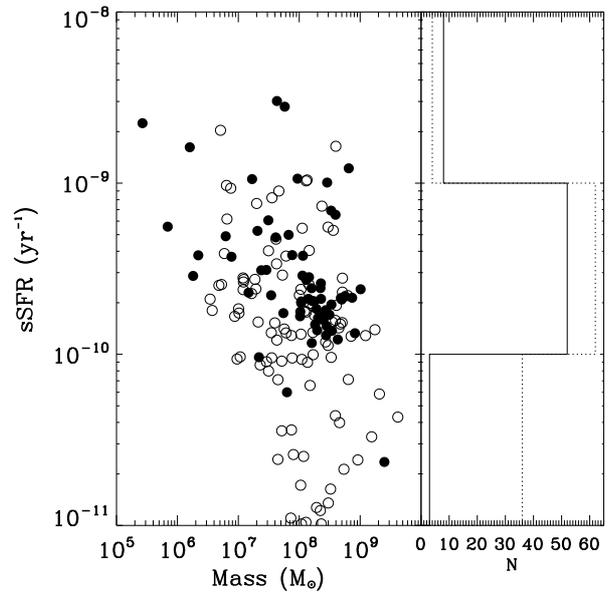}
\caption{sSFR versus mass for Sknot (filled circles) and knots in Sknot+diffuse and Mknots galaxies (open circles). To the right, the distribution function (solid line is for Sknot galaxies and dotted line is for knots in Sknot+diffuse and Mknots galaxies).}
\label{surface}
\end{figure}

In our sample we detected two kinds of star-forming knots. One are classified as Sknot galaxies, where there is one star-forming knot, and it is not possible to detect diffuse emission. The other are knots which belong to galaxies with one or more knots of SF and diffuse emission. We detected similitudes and differences in the properties of both classes.  Fig. \ref{surface} represents the distribution of sSFR of both populations, showing a maximum at the same sSFR value. 1/SSFR can also be used as an estimation of age as the time scale to form the stars we see now if they would have been formed at the present SFR. With this idea, the typical time required to form both systems - starbursts spread along the disks or centered isolated ones - is around 2$\times$10$^9$ years. Knots in Sknot+diffuse and Mknots galaxies do show, however, a wider range of values, with some cases that can require a formation time of up to 10$^{11}$ years. This can be, however, a caveat of our determination of parameters, as we have explained in section \ref{mass_sfr}. 

\begin{figure*}[t]
\centering
\includegraphics[width=18.0cm]{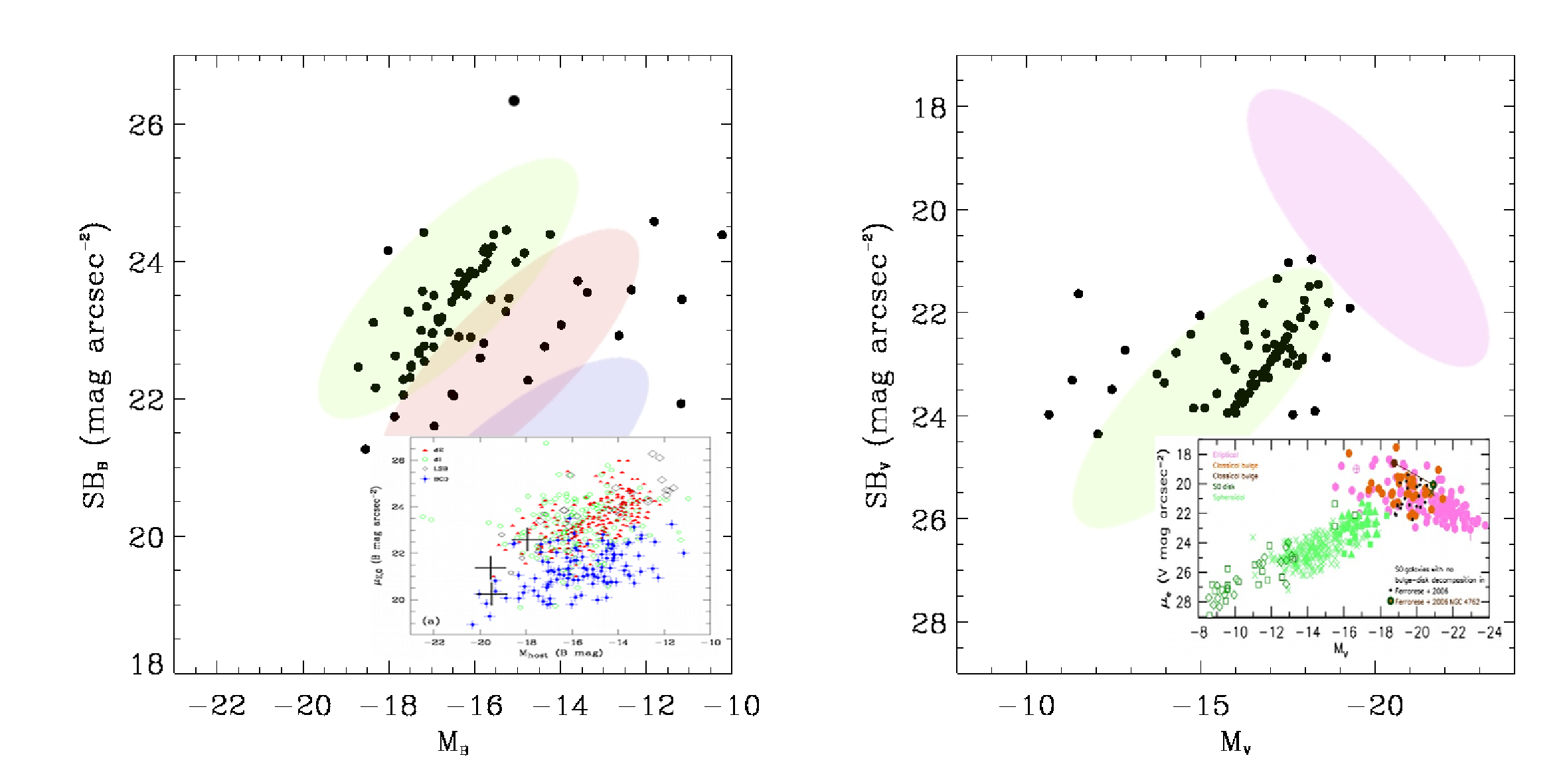}
\caption{Surface brightness versus absolute magnitude in the $B-$band (left panel) and $V-$band (right panel) for Sknot galaxies. Figures are adapted from \citet[][their Fig. 9]{amorin2012} for $B-$band and \citet[][Their Fig. 7]{kormendy2012} for $V-$band (the inset figure in each panel). Elliptical regions of different colors are overplotted to show the different class of galaxies, following the \citet{amorin2012} and \citet{kormendy2012} diagrams.}
\label{amorin_kormendy}
\end{figure*}

For both classes we measured the surface brightness in the F814W-band. Sknot have lower values, characteristic of low surface-brightness galaxies. For Sknot galaxies we measured the surface brightness and absolute magnitude in $B-$ and $V-$band, to compare with results in the literature. In particular, we compare the photometric values of Sknot galaxies with those provided in \citet[][Fig. 9]{amorin2012} and \citet[][Fig. 7]{kormendy2012}, respectively. In Fig. \ref{amorin_kormendy}a we show the surface brightness in $B-$band versus $B-$band absolute magnitude as in \cite{amorin2012}, and Fig. \ref{amorin_kormendy}b shows the surface brightness in $V-$band versus $V-$band absolute magnitude as in \cite{kormendy2012}. At the right upper corner of each panel, and with the same scale, we show the data points of this work. In the figures we overplotted a shadowed elliptical region where most of our points are in the plots of \cite{amorin2012} and \cite{kormendy2012}. A direct comparison suggests that Sknot galaxies fall in the parameter space populated by spheroidals and ellipticals \citep{kormendy2012}, and dwarf ellipticals and dwarf irregulars galaxies, as measured by \cite{papaderos2008} and compiled by \cite{amorin2012}. 

\section{Conclusions}
\label{sec:results}

We have selected a sample of young starburst galaxies in the COSMOS field using a new tailor-made color-color diagnostic. The final catalogue consists of 220 galaxies with EW$>$80 \AA\ in both H$\alpha$ and [OIII]. The sample has been identified by using both the spectra from the zCOSMOS catalogue in the redshift range 0.1 $\le z \le$ 0.48, and the photometric redshift catalogue using the SUBARU intermediate band filters in the redshift ranges 0.007 $\le z \le$ 0.074, 0.124 $\le z \le$ 0.177 and 0.23 $\le z \le$ 0.274.

In order to characterize the morphology of the star-forming regions in our sample galaxies we perform an isophotal analysis of the HST/ACS high spatial resolution images. From this analysis we classify the starburst galaxies in COSMOS ($z<0.5$) as Sknot galaxies, Sknot+diffuse galaxies, and Mknots galaxies, whether they consist of a single knot of star formation, a single knot surrounded by diffuse emission, or several knots of star formation, respectively. The stellar masses of the knots and the galaxies were calculated using photometric and spectroscopic data from the COSMOS database. The mean and maximum mass of the starburst galaxies are 10$^{8.9}$ M$_\sun$ and 10$^{11}$ M$_\sun$. We have compared this distribution with that obtained for the whole sample of galaxies in COSMOS at the same redshift range. The resulting mean mass is 10$^9$ with a (similar) lack of galaxies with $M/M_{\odot} >$ 10$^{10}$. Therefore, the mass distribution of the starburst galaxies follows the same distribution as the whole sample of galaxies in COSMOS.

The masses for individual knots in Sknot+diffuse and Mknots galaxies vary with the distance to the center of the galaxy, increasing their mass the closer to the center they are. Masses of knots are typically one order of magnitude below that of the host galaxy, peaking at 10$^{7.7}$M$_\odot$ (see Fig. 12). The specific characteristics of the starburst knots as a function of their distance to the center of their host galaxy are similar. The surface SFR and surface mass do not show any footprint of their particular location in the galaxy. 

The SFR of the knots follows the same trend with their mass to that of their host galaxy, with an offset to higher values of the SFR of the knots of about a factor 3. This is a tendency similar to that found by \citet{guo2012} for higher z galaxies.

Even though the observational criteria to define clumps differs from different authors, and no direct comparison with other published papers can be done, we have made the exercise to define a "clumpy fraction". This parameter, defined in section \ref{comp_lit}, gives a value of f$_{clumpy}$=0.24, which is higher than other studies \citep[e.g.][f$_{clumpy}$=0.08]{Murata2014}. The reasons for this difference might be in the different mass ranges of the samples, and the precise criteria to identify starburst galaxies and star-formation clumps, as we showed in section \ref{comp_lit}. 

The fraction of clumpy galaxies is a prediction of the numerical simulations. It is expected to first increase with redshift until $z>4$ where simulations predict less clumpy and more compact star-forming galaxies than at lower redshifts \citep{ceverino2015}. This is consistent with UV observations of bright clumps in high-$z$ galaxies \citep{guo2015}, where they found the fraction of clumpy galaxies 0.6 with stellar masses of log(M/M$_{\odot}$)=9-10 in the redshift range z = 0.5-3 (see also \citet{Elmegreen2007} and \citet{tadaki2014}). From this work the clumpy fraction drops down to 0.24 at $z<0.5$, for the starbursts in COSMOS, which have a typical stellar mass of $10^{8.9}~{\rm M}_{\odot}$. Our result therefore would also agree with the expected trend in the fraction of clumpy galaxies with $z$ predicted by the numerical simulations.

Different properties of the knots, with respect to the absolute $B-$band magnitude of their host galaxies in our sample were compared with local and high-redshift galaxies \citep{Elmegreen2013}. We found that the clump surface density and clump mass versus host galaxy {\it B}-band absolute magnitude have a slope between those for the local and high-redshift sample, implying that for a given absolute $B-$band magnitude of the galaxy, the mass and surface density of the knots of the sample of this paper have higher values than those of clumps in local spirals, and lower than those found in high-redshift massive star-forming regions.

The L(H$\alpha$) versus size of star-forming regions have been investigated by several authors, searching for an universal relation. Taking the benefit of the excellent spatial resolution provided by the HST images, we have built the L(H$\alpha$) versus size for all individual knots of the 220 galaxies of our catalogue. We obtain a slope of 2.48 $\pm$ 0.05. This is a value similar to 2.5 obtained by \citet{Fuentes-Masip2000} for local, resolved giant HII regions measured with the Fabry-Perot technique. Note, however, that this value differs from the 2.78 slope given by \citet{Wisnioski2012}, in which the authors include high-redshift galaxies. The spatial resolution in \citet{Fuentes-Masip2000} and this work allow us to have a good confidence in the results. For high-redshift galaxies the resolution is worse and can include errors in the measurements, which cannot be quantified.

Sknot galaxies ($\sim$38$\%$) show photometric structural properties that differ from star-forming knots in the other classes of galaxies (Sknot+diffuse and Mknots). Sknot galaxies have lower surface-brightness and lie in the dwarf spheroidal, dwarf irregular and elliptical regions in the surface-brightness versus absolute magnitude ($M_B$ and $M_V$) diagrams. The possibility of an evolutionary trend among different dwarf systems was already proposed by \citet{papaderos1996}, and we suggest that Sknot galaxies in COSMOS may be examples of a transitional phase between BCD starbursts and dwarf spheroidals.

\begin{acknowledgements}
This work has been funded by the Spanish MINECO, Grant ESTALLIDOS, AYA2013-47742-C4-2P and AYA2010-21887-C04-04. JMA acknowledges support from the European Research Council Starting Grant SEDMorph (P.I. V. Wild).
Rodrigo HG acknowledges the FPI grant from MINECO within ESTALLIDOS project. We thank the referee for his/her constructive comments which helped to improve the paper. Thanks to Ismael Mart\'inez Delgado for his help with the installation and use of FOCAS software. Thanks to Mercedes Filho for her comments and corrections to the text.
\end{acknowledgements}

\bibliographystyle{aa}
\bibliography{biblio.bib}

\end{document}